\documentclass[9pt,a4paper,twocolumn,fleqn]{extarticle}

\usepackage[T1]{fontenc}
\usepackage{lmodern}
\usepackage[top=17mm,bottom=19mm,left=15mm,right=15mm]{geometry}
\usepackage[authoryear,round]{natbib}
\usepackage{graphicx}
\usepackage{subcaption}
\usepackage{amsmath,amssymb}
\usepackage{placeins}
\usepackage{caption}
\usepackage{xcolor}
\colorlet{blue}{black}
\colorlet{red}{black}
\usepackage[hidelinks]{hyperref}
\usepackage{microtype}
\usepackage{dblfloatfix}
\hypersetup{
  pdftitle={BRISK: An accelerated wide-field imager for radio astronomy},
  pdfauthor={Oliver Ogden and Wesley Armour},
  pdfsubject={Wide-field radio interferometric imaging}
}

\DeclareRobustCommand{\VAN}[3]{#2}
\let\VANthebibliography\thebibliography
\def\thebibliography{\DeclareRobustCommand{\VAN}[3]{##3}\VANthebibliography}
\providecommand{\degr}{^{\circ}}
\providecommand{\la}{\lesssim}
\begin{document}
\twocolumn[
\begin{minipage}{\textwidth}
\begin{center}
{\LARGE\bfseries BRISK: An accelerated wide-field imager for radio astronomy\par}
\vspace{0.7em}
{\large Oliver Ogden \quad Wesley Armour\textsuperscript{*}\par}
\vspace{0.3em}
{\small Oxford E-Research Centre, 7 Keble Rd, Oxford, OX1 3QG, UK\par}
{\footnotesize \textsuperscript{*}E-mail: wes.armour@oerc.ox.ac.uk\par}
\end{center}
\vspace{0.3em}
\begin{abstract}
Wide-field interferometric imaging with non-coplanar arrays is computationally expensive, particularly as we move towards imaging large datasets from next-generation telescopes such as the Square Kilometre Array (SKA). The dominant imaging cost arises during the gridding and degridding steps, where large $w$-terms require correction. In $w$-projection, this correction uses $w$-dependent convolution kernels whose support grows rapidly with field of view (FoV) and maximum $|w|$. In $w$-stacking, visibilities are partitioned into discrete $w$-layers, each requiring a 2D Fourier transform and image-domain phase correction; the required number of layers increases with $|w|_{\max}$ and FoV. We present BRISK (\textbf{B}aseline-aware \textbf{R}ank-compressed \textbf{I}nterferometric \textbf{S}ky-imaging \textbf{T}oolkit), a wide-field interferometric imaging algorithm that accelerates $w$-projection by exploiting the low-rank structure of its convolution kernels. For each $w$-plane, the full 2D gridding kernel is compressed using a truncated randomised singular value decomposition (SVD), yielding a compact representation from which the SVD-compressed kernel can be reconstructed on the fly while preserving the dominant kernel energy. We benchmark memory usage, runtime, and accuracy against \textsc{casa}'s $w$-projection and \textsc{wsclean}'s $w$-stacking, $w$-gridding and image-domain gridding algorithms on a simulated $15 \times 15$ grid of point sources, as well as the SKA1-Low GLEAM simulated measurement sets.  BRISK runs up to {\color{blue} 2443 times} faster than \textsc{casa}, and {\color{blue}up to 132 times faster than \textsc{wsclean}}, while maintaining a peak-flux bias of less than 1\% at the largest image sizes and comparable peak memory usage.
\end{abstract}
\vspace{0.8em}
\end{minipage}
]

\section{Introduction}
In radio interferometric imaging, the sky brightness distribution can be reconstructed from the sampled visibilities collected by a telescope array \citep{ThompsonMoranSwenson2017}. For small fields of view and coplanar arrays, the relationship between the visibilities and the sky is a two-dimensional Fourier transform pair. For wide-field observations, or when pointing far from zenith with non-coplanar arrays, the third baseline coordinate $w$ introduces a direction-dependent phase error that must be corrected to preserve image accuracy.

Two algorithms dominate modern wide-field imaging: $w$-projection \citep{Cornwell2008} and $w$-stacking \citep{HumphreysCornwell2011, Offringa2014}. In $w$-projection, the $w$-term is corrected during gridding by convolving each visibility with a $w$-dependent kernel in the Fourier domain. This is accurate, but the kernel support grows with $|w|$ and field of view (FoV), driving up both the compute cost of gridding and the memory required to store the kernel bank. In $w$-stacking, the three-dimensional problem is split into a set of $w$-layers, each of which is Fourier-transformed and phase-corrected independently before being combined in the image plane. This eliminates wide-support kernels at the cost of many large Fourier transforms, with the number of layers growing with $|w|$ and image resolution.

Other approaches include $w$-gridding, image-plane faceting \citep{CornwellPerley1992}, full three-dimensional Fourier transforms \citep{Perley1999}, hybrid snapshot schemes such as $w$-snapshots \citep{CornwellVoronkovHumphreys2012}, and image-domain gridding, which replaces oversampled convolution kernels with small sub-grid Fourier transforms \citep{vanderTol2018}. Hybrid $w$-stacking/$w$-projection methods have also been developed \citep{Pratley2019, Pratley2020}. Despite this progress, $w$-correction remains a dominant computational cost in imaging, and that cost is set to grow sharply with the next generation of interferometers.

Instruments such as the Square Kilometre Array (SKA) will produce volumes of visibility data far beyond anything handled by current imagers, sampled over wide fields of view and at high angular resolution \citep{Scaife2020, Braun2019}. Under these conditions, the computational cost of wide-field imaging becomes a critical bottleneck in existing imaging software. \textsc{casa}  \citep{CASATeam2022} implements an efficient $w$-projection algorithm during the gridding step and only requires a single Fourier transform to form the dirty image. This approach results in relatively modest run-time memory requirements, but the computational cost of applying the wide support convolution gridding kernels leads to long runtimes for large images. In addition, the memory storage required to hold the kernel look-up tables can be significant for datasets with large $w$-values. \textsc{wsclean} \citep{Offringa2014} takes the opposite trade-off, using $w$-stacking to reduce per-visibility gridding cost but holding many concurrent $w$-layers in memory, which becomes prohibitive at high resolution. \textsc{wsclean} also provides alternative gridders: $w$-gridding \citep{Ye2022, Arras2021}, which uses a kernel in the $w$-direction as well as in $u$ and $v$ and achieves high accuracy at moderate memory cost, and image-domain gridding (\textsc{idg})\citep{vanderTol2018}. Both are accurate, and at large image sizes, benefit from parallelised deconvolution to keep total runtimes tractable. For SKA-scale imaging, where images routinely contain tens of thousands of pixels per dimension, runtime and memory footprint become equally critical constraints.

{\color{red}\textsc{wsclean} has the option to run \textsc{idg} on the CPU, GPU and a hybrid of both. This comes with the benefit that GPU gridders achieve high per-visibility throughput \citep{Veenboer2020, Bhatnagar2022}. However, at large image sizes, the image-domain arrays, Fourier transforms, and deconvolution set both the memory footprint and much of the runtime. In Pigi, device memory limits Stokes-I images to $48{,}000^2$ pixels on a 40-GB A100 and $64{,}000^2$ pixels on a 64-GB MI250X \citep{Hodgson2026}, and hybrid schemes that keep the grid in host memory \citep{Veenboer2020} accelerate only the gridding step, leaving the deconvolution on the CPU. The approach taken in BRISK is CPU-first rather than CPU-only: it reduces the memory traffic of the gridding kernel, which is a limiting resource on either architecture, and it is implemented for large-memory CPU nodes on which images of side lengths up to 131,072 pixels can currently be held.}
{\color{red}A same-hardware-class, stage-by-stage comparison of BRISK with the CPU, GPU and hybrid \textsc{idg} modes is provided in Appendix~\ref{app:idg_gpu_breakdown}.}

Previous work has explored alternate approaches to simplifying the $w$-projection kernels. Radially symmetric representations have been formulated using the one-dimensional Hankel transform \citep{Pratley2019}, reducing the kernel construction cost by orders of magnitude but leaving the per-visibility 2D convolution during gridding unchanged. \citet{Merry2016} showed that $w$-projection kernels can be represented closely as one-dimensional separable kernels, with an approximation error that scales as the fourth power of FoV; this permits 1D convolution during gridding but loses accuracy at wide FoV unless combined with faceting or snapshots.

In this work, we introduce BRISK, a wide-field imaging pipeline that accelerates $w$-projection by exploiting the low-rank structure of the gridding convolution function itself. For each $w$-plane, a truncated randomised SVD \citep{HalkoMartinssonTropp2011} decomposes the two-dimensional kernel into a compact set of separable rank-1 basis vectors. {\color{blue}Unlike the single-mode approximation of \citet{Merry2016}, BRISK retains as many separable modes as the kernel requires to meet a chosen retained-energy threshold at each $w$ and FoV. The radial Hankel-transform method of \citet{Pratley2019} instead reduces kernel-construction cost while retaining a two-dimensional kernel during gridding.} The representation is small enough that any required part of any kernel can be reconstructed on the fly during gridding, giving efficient 2D convolution, significantly reduced kernel storage, and lower memory bandwidth during gridding and degridding.

The remainder of the paper is structured as follows. Section~\ref{sec:wproj} reviews the $w$-projection algorithm. Section~\ref{sec:SVDkernels} describes the SVD decomposition of the gridding kernel and its use during gridding and predict. Section~\ref{sec:imager} sets out the full BRISK imaging and deconvolution pipeline. Section~\ref{sec:results} compares the performance and accuracy of the new pipeline against \textsc{wsclean} and \textsc{casa}, culminating with the conclusions in Section \ref{sec:conclusion}.

\section{The $w$-Projection Algorithm}
\label{sec:wproj}
This section will detail the $w$-projection algorithm mathematically as a method for correcting for the $w$-component in wide-field imaging. The relationship between the sampled visibilities $V(u,v,w)$ and the sky brightness distribution $I(l,m)$ is given via the imaging equation as follows:
\begin{equation} \label{equation:imagingintegral}
\begin{aligned}
V(u,v,w) ={}& \int\!\!\int \frac{I(l,m)}{\sqrt{1-l^2-m^2}} \\
&\quad \times e^{-2\pi i(ul + vm + w(\sqrt{1-l^2-m^2}-1))}\,dl\,dm
\end{aligned}
\end{equation}
{\color{blue}where} $(u,v,w)$ are the baseline coordinates measured in wavelengths and $(l,m)$ are the direction cosines on the celestial sphere. For the purpose of this research, only the $XX$ polarisation will be imaged; any primary beam effects are ignored, and visibilities are already calibrated.

{\color{red}
For small FoVs, $l$ and $m$ are small enough that $\sqrt{1-l^2-m^2}\approx1$, and equation~\ref{equation:imagingintegral} reduces to a two-dimensional Fourier transform; the same holds for observations near zenith, where $w$ itself is negligible. For wide-field observations with non-coplanar baselines, however, the $w$-dependent phase term is no longer negligible:
\begin{equation}
\label{equation:wphase}
e^{-2\pi i w(\sqrt{1-l^2-m^2}-1)}.
\end{equation}
This term introduces a direction-dependent distortion that must be corrected during imaging.
}

Using the Fourier convolution theorem, the imaging equation can be rewritten such that each visibility is convolved in the Fourier domain with a $w$-dependent convolution kernel. The image space representation of this kernel is represented via the following phase screen:
\begin{equation}\label{equation:phasescreen}
G_w(l,m) =
\frac{1}{\sqrt{1-l^2-m^2}}
e^{-2\pi i w(\sqrt{1-l^2-m^2}-1)}
\end{equation}

This then allows any visibility with non-zero $w$ to be calculated from the convolution between the visibility with zero $w$-component and the known function $\tilde{G}(u,v,w)$, which is the Fourier transform of the phase screen in equation \ref{equation:phasescreen}. Instead of generating a convolution kernel for every single visibility, the $w$-range of the data set is discretised into a set of $w$-planes. A kernel is then precomputed for each and stored in a kernel bank.

In practice, convolutional gridding also makes use of an anti-aliasing (AA) kernel to limit the spectral leakage caused by the discretisation of the continuous visibility space onto a regular grid. Each visibility is convolved with an AA kernel in the Fourier domain, and the kernel support and type are chosen to minimise aliasing while maintaining accuracy. Common choices include the prolate spheroidal wave function (as used in \textsc{casa}) \citep{Schwab1984} and the Kaiser-Bessel function (as used in \textsc{wsclean}) \citep{Jackson1991, Offringa2014}. Typical implementations of the AA kernel use a support of between 5 and 9 pixels, depending on the chosen function and oversampling factor. In line with \textsc{wsclean}, our pipeline automatically selects a support of 7 pixels as this is a good balance between compute time and gridding accuracy \citep{Offringa2014}.

Within the framework of $w$-projection, the AA kernel is incorporated directly with the $w$-projection kernel to give the Gridding Convolution Function (GCF). The effective image domain of this kernel is the product of the anti-aliasing function $A(l,m)$ and the $w$-dependent phase screen $G_w(l,m)$,
\begin{equation}
K_w(l,m) = A(l,m)\,G_w(l,m)
\end{equation}
The GCF used during gridding is then obtained through the Fourier transform of this product,
\begin{equation}
\tilde{K}_w(u,v) =
\mathcal{F}\{A(l,m)G_w(l,m)\}.
\end{equation}
The support of the GCF is dominated by the $w$-projection kernel and therefore grows with both the magnitude of $w$ and the FoV of the observation. With a larger $w$, the phase term in $G_w(l,m)$ becomes increasingly oscillatory in the image plane, hence resulting in the GCF needing to be evaluated over a larger support. This increase in kernel size increases both the memory footprint of the precomputed kernel bank and the number of arithmetic operations required during the gridding and predict stages. For wide-field observations with large FoV and high $w_{\max}$, these effects can significantly increase computational cost and memory requirements of the imaging pipeline, therefore motivating the development of methods that reduce the cost of the GCF.

\section{Implementation of SVD in w-projection}
\label{sec:SVDkernels}

\begin{figure*}
    \centering

    \begin{subfigure}[b]{0.48\textwidth}
        \centering
        \includegraphics[width=\textwidth]{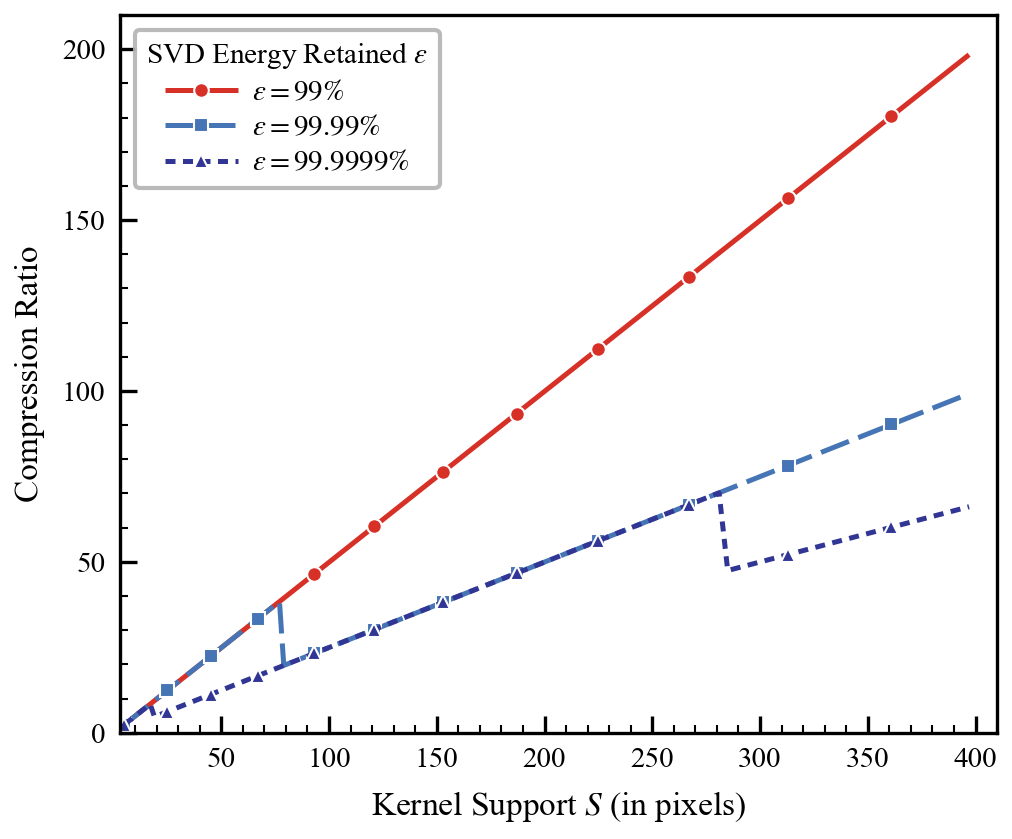}
        \caption{Individual compression ratio ($C_{\mathrm{SVD}}$) per $w$-plane as a function of kernel support $S$, for three retained-energy thresholds $\epsilon$.}
        \label{fig:compratio_vs_support}
    \end{subfigure}
    \hfill
    \begin{subfigure}[b]{0.48\textwidth}
        \centering
        \includegraphics[width=\textwidth]{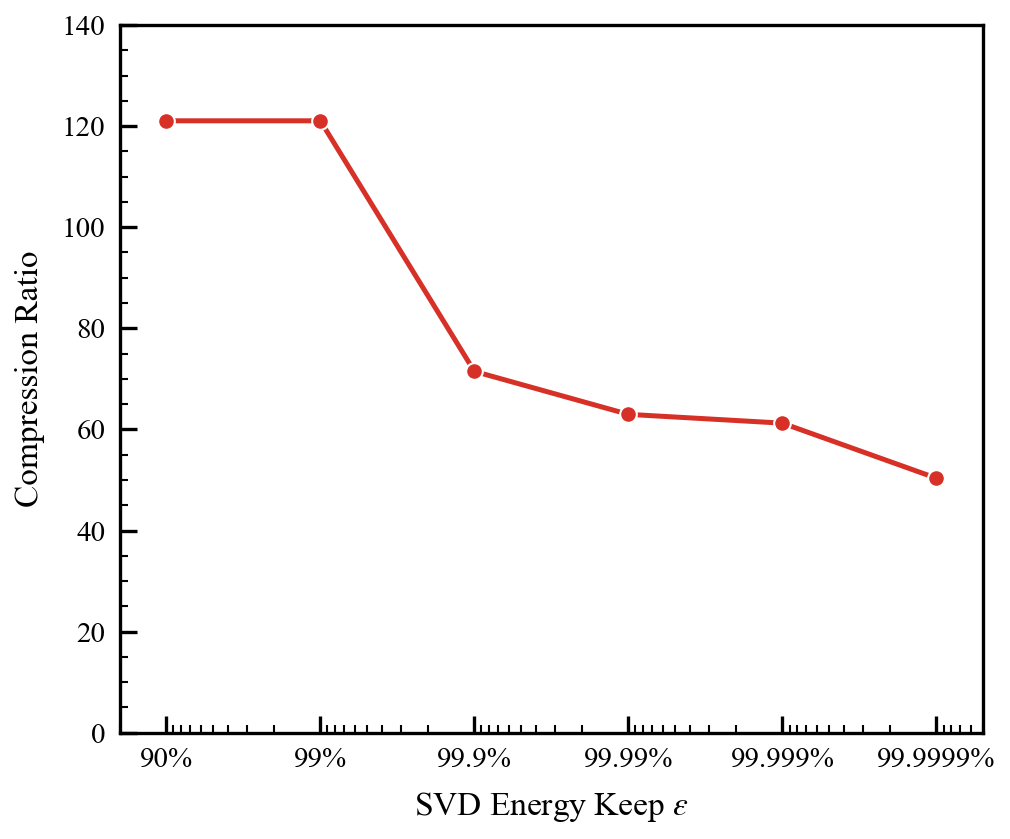}
        \caption{Compression ratio of the full kernel bank versus retained-energy threshold $\epsilon$.}
        \label{fig:overallcompratio}
    \end{subfigure}

    \caption{Memory reduction from storing the truncated SVD representation of the GCF rather than the full two-dimensional kernel for every $w$-plane.}
    \label{fig:twographs}
\end{figure*}

This section describes the implementation of Singular Value Decomposition (SVD) compression within the $w$-projection imaging pipeline, including construction of the compressed GCF kernel bank, on-the-fly kernel reconstruction during gridding, visibility prediction (degridding), fractional-pixel oversampling, and interpolation between neighbouring $w$-planes.

{\color{blue}
\subsection{Singular Value Decomposition of the Gridding Kernel}

For any given $w$-plane, the GCF can be sampled on a discrete grid and represented as a matrix $K_w\in\mathbb{C}^{N\times N}$, where $N$ is the full width of the convolution kernel in pixels. In this work, the support parameter $S$ denotes the kernel radius, such that
\begin{equation}
N=2S+1.
\end{equation}

Applying the SVD to the sampled kernel gives the expanded form
\begin{equation}
K_w(u,v)=
\sum_{i=1}^{N}
\sigma_i\,u_i(u)\,v_i^*(v),
\end{equation}
where the real, non-negative singular values $\sigma_i$ are ordered from largest to smallest, and $u_i$ and $v_i$ are the corresponding left and right singular vectors. Each term in this expansion is separable in $u$ and $v$, so the two-dimensional kernel can be represented as a sum of rank-1 components.

The $w$-projection kernels are amenable to low-rank factorisation because the image-domain phase screen is a smooth, slowly varying function that depends only on the direction-cosine distance from the phase centre. After multiplication by the AA kernel, which is typically implemented as a separable rank-1 function, and Fourier transformation, the resulting GCF remains dominated by a small number of smooth separable components and is therefore well represented by a low-rank truncated expansion. In the narrow-field limit,
\begin{equation}
G_w(l,m) \approx e^{-\pi i w(l^2+m^2)},
\end{equation}
which is exactly separable in $l$ and $m$ and therefore produces a rank-1 kernel. At wider fields of view, the higher-order terms neglected by this approximation introduce additional structure into the GCF, causing its energy to be distributed across more singular modes. The required rank therefore increases as the kernel departs from the narrow-field limit, but remains much smaller than the full kernel representation.

The low-rank approximation is obtained by retaining only the $r_{\rm SVD}$ largest singular values, where $r_{\rm SVD} \ll N$ is the smallest integer satisfying the Frobenius-energy criterion
\begin{equation}
\frac{\sum_{i=1}^{r_{\rm SVD}} \sigma_i^2}
     {\sum_{i=1}^{N} \sigma_i^2}
\ge \epsilon ,
\end{equation}
where $\epsilon$ is the retained fraction of the total kernel energy. The resulting approximation is
\begin{equation}
K_w(u,v)\approx
\sum_{i=1}^{r_{\rm SVD}}
\sigma_i\,u_i(u)\,v_i^*(v).
\end{equation}
The retained-energy threshold $\epsilon$ is distinct from the spatial support threshold $\kappa$ introduced in Section~\ref{sec:imager}. The spatial threshold is applied first and determines how far the GCF extends in the $uv$-plane before decomposition. The spectral threshold is then applied to the cropped kernel and determines how many of its singular modes are retained. The two thresholds therefore control different approximations and may be selected independently.

For each $w$-plane, storing the full two-dimensional kernel requires $N^2$ complex values. The truncated representation instead stores $r_{\rm SVD}$ pairs of one-dimensional singular vectors of length $N$, together with $r_{\rm SVD}$ real singular values. Its dominant storage requirement is therefore approximately $2Nr_{\rm SVD}$ complex values, giving a compression ratio
\begin{equation}
C_{\rm SVD}
\simeq
\frac{N^2}{2Nr_{\rm SVD}}
=
\frac{N}{2r_{\rm SVD}}
=
\frac{2S+1}{2r_{\rm SVD}}.
\end{equation}
The comparatively small storage required for the singular values themselves has been neglected in this expression.

This relation also makes clear that the cost of the compressed representation is controlled jointly by the kernel support width $N$ and the retained rank $r_{\rm SVD}$. Even if the rank remained fixed, the storage and basis-loading costs would increase linearly as the kernel support grows. The retained rank is itself determined by the singular-value spectrum of the sampled kernel and is therefore not a function of $w$ alone. Increasing $|w|$ or field of view generally widens the kernel support and introduces additional structure, while changes to the sampling or to the terms incorporated into the GCF may also alter the number of modes required to satisfy a fixed value of $\epsilon$.

This work reduces the computational cost and memory footprint of $w$-term corrections during gridding and degridding. While it may be possible to extend this method to include aperture terms, direction-dependent effects, and wide-band sky modelling, this has been left to future work, as discussed in Section~\ref{sec:conclusion}.

The leading modes are computed using the randomised SVD of \citet{HalkoMartinssonTropp2011}, avoiding calculation of the complete dense decomposition when only $r_{\rm SVD}\ll N$ modes are required.

Figures~\ref{fig:compratio_vs_support} and \ref{fig:overallcompratio} show the resulting memory reduction. The per-plane compression ratio increases with support width, so the widest kernels obtain the greatest benefit. Tightening $\epsilon$ reduces the compression ratio by retaining additional modes, but the complete kernel bank remains compressed by approximately $50\times$ at $\epsilon=99.9999\%$.

Figure~\ref{fig:cumnrg_vs_singularvalues} shows the cumulative energy spectrum for kernels spanning $w=0$ to $w=10^6$. For $w\leq10^3$, the first singular value contains more than 99.99\% of the kernel energy. Four modes are required at $w=10^4$, eight at $w=10^5$, and 29 at $w=10^6$. These values span regimes encountered by current and next-generation arrays: international LOFAR reaches characteristic values of $w\sim10^6$, while SKA-Low and SKA-Mid span approximately $w\sim10^4$--$10^5$ and $w\sim10^5$--$10^6$, respectively, with still larger values possible in the upper SKA-Mid bands \citep{vanHaarlem2013,SKALowSpecs,SKAMidSpecs}. Even across this range, the GCFs remain compressible using at most a few tens of retained modes.

\begin{figure}
    \centering
    \includegraphics[width=0.98\columnwidth]{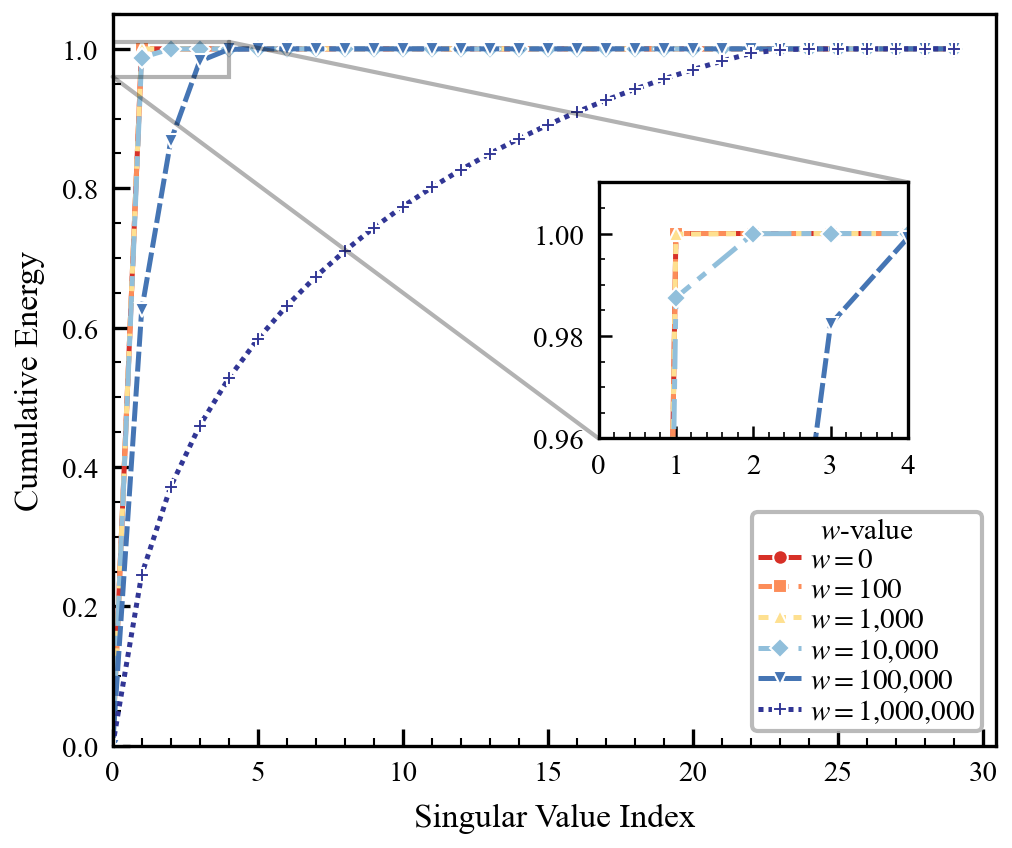}
    \caption{Cumulative energy as a function of retained singular-value index for kernels spanning six orders of magnitude in $w$. Low-$w$ kernels are effectively rank-1, while $w=10^6$ requires $\sim29$ singular values for full reconstruction; the inset separates the overlapping low-$w$ curves.}
    \label{fig:cumnrg_vs_singularvalues}
\end{figure}
}

{\color{blue}
\subsection{On-the-fly Kernel Reconstruction During Gridding}
\label{sec:kernel_recon}

The computational advantage of the SVD representation becomes apparent during gridding. Conventional $w$-projection streams up to $N^2$ complex kernel coefficients for each visibility and uses them to update the corresponding grid cells. For large kernel supports, the resulting gridding operation can therefore be limited by off-chip memory bandwidth.

BRISK instead stores the kernel as $r_{\rm SVD}$ pairs of one-dimensional factors and partitions the $uv$-grid into non-overlapping tiles. For a visibility $V_j$ intersecting a tile over a region of size $T_u\times T_v$, the contribution to each tile-local grid cell is evaluated directly as
\begin{equation}
\label{equation:tiled_SVD}
G(p,q)
\mathrel{+}=
V_j
\sum_{i=1}^{r_{\rm SVD}}
\sigma_i\,
\tilde{u}_i(p)\,
\tilde{v}_i^{*}(q),
\end{equation}
where $(p,q)$ are tile-local coordinates and $\tilde{u}_i$ and $\tilde{v}_i$ contain only the portions of the singular vectors intersecting the tile. The two-dimensional kernel patch is not materialised. Evaluating a $T_u\times T_v$ region requires $r_{\rm SVD}(T_u+T_v)$ complex factor entries, rather than the $T_uT_v$ coefficients of a stored two-dimensional patch.

Before gridding, each visibility is assigned to every tile touched by its kernel footprint. Each tile then evaluates only the intersecting part of that footprint and writes exclusively to its own grid cells, allowing different tiles to be processed concurrently without locks or private copies of the full grid.

The use of non-overlapping grid regions with exclusive write ownership is closely related to the multithreaded gridding scheme described by \citet{Golap2015} and implemented in \textsc{casa}. In both approaches, each thread updates only the part of a visibility footprint that intersects its assigned region, avoiding locks or private copies of the full grid. However, in BRISK, rather than retrieving values from a stored two-dimensional kernel, the tile loads the required subsets of the one-dimensional SVD factors and applies them directly to its local accumulator.

Within each tile, visibilities are grouped by $w$-plane index and oversampling bin (Section~\ref{sec:oversampling}) so that the same basis vectors can be reused across consecutive reconstructions. Equation~\ref{equation:tiled_SVD} is a length-$r_{\rm SVD}$ dot product consisting primarily of multiply-accumulate operations, which map efficiently onto SIMD vector instructions such as AVX2 and AVX-512 and allow several kernel coefficients to be reconstructed simultaneously within a single CPU operation.

{\color{red}For a tile-local footprint of size $T_u\times T_v$, retained rank $r_{\rm SVD}$, and scalar size $b$ bytes, data-only gridding performs
\begin{equation}
F_{\rm BRISK}
=
r_{\rm SVD}
\left(
8T_uT_v+6T_u
\right)
\end{equation}
floating-point operations. Loading each required subset of the complex $u$ and $v$ factors once requires at most
\begin{equation}
Q_{\rm BRISK}
=
2b\,r_{\rm SVD}(T_u+T_v)
\end{equation}
bytes. The resulting kernel-stream arithmetic intensity is therefore
\begin{equation}
I_{\rm BRISK}
=
\frac{8T_uT_v+6T_u}
     {2b(T_u+T_v)}
\quad
\mathrm{FLOP\,byte^{-1}}.
\end{equation}
The retained rank cancels because each additional mode increases the work and factor traffic proportionally. For an unsplit square footprint of width $N$, this reduces to
\begin{equation}
I_{\rm BRISK}
=
\begin{cases}
N/4+0.1875, & \text{double precision},\\
N/2+0.375,  & \text{single precision}.
\end{cases}
\end{equation}
Tile boundaries reduce these ideal values because the required one-dimensional factor subsets must be revisited in each intersected tile.

For comparison, the \textsc{casa} 6.5 \textsc{WProject} loop loads one 8-byte complex single-precision kernel value per grid cell, performs one complex multiply-add, and updates the normalization sum. Assuming that the grid remains cache-resident, its optimistic kernel-stream intensity is
\begin{equation}
I_{\text{\textsc{casa}}}
=
\frac{9}{8}
=
1.125\ \mathrm{FLOP\,byte^{-1}}.
\end{equation}
Including grid reads and writes reduces this value to $0.375\ \mathrm{FLOP\,byte^{-1}}$.

Applying this model gives the arithmetic intensities shown in Table~\ref{tab:arithmetic-intensity}. When the dirty-image and PSF calculations are fused, the same SVD-factor loads are reused and the intensity at $2048^2$ pixels increases to $12.13$~FLOP~byte$^{-1}$.

\begin{table}
\color{red}
\centering
\caption{\color{red} Source-level arithmetic intensity (FLOP byte$^{-1}$) of the production BRISK gridding loop, compared with the optimistic kernel-only value for the \textsc{casa} 6.5 \textsc{WProject} implementation. Gain is relative to double-precision BRISK.}
\label{tab:arithmetic-intensity}
\setlength{\tabcolsep}{5pt}
\resizebox{\columnwidth}{!}{%
\begin{tabular}{rcccc}
\hline\hline
Image size
& \multicolumn{1}{c}{BRISK (double)}
& \multicolumn{1}{c}{BRISK (single)}
& \multicolumn{1}{c}{\textsc{casa}}
& \multicolumn{1}{c}{Gain} \\
\hline
$1024$ & 4.85 &  9.71 & 1.125 & $4.31\times$ \\
$2048$ & 6.13 & 12.25 & 1.125 & $5.45\times$ \\
$4096$ & 6.81 & 13.63 & 1.125 & $6.06\times$ \\
$8192$ & 9.14 & 18.28 & 1.125 & $8.13\times$ \\
\hline
\end{tabular}%
}
\end{table}

These values are source-level kernel- and factor-stream intensities rather than measurements from hardware performance counters. The BRISK calculation conservatively treats every tile-local factor subset as a cold load, so reuse between visibilities can only increase its intensity. Conversely, the optimistic \textsc{casa} comparison excludes grid traffic. Common costs such as grid initialisation, final tile writeback, visibility metadata, and the FFT are excluded from both kernel-stream models.

For $w<0$, the GCF is the complex conjugate of the corresponding positive-$w$ kernel. BRISK therefore stores only non-negative $w$-planes and changes the sign of the imaginary contribution when applying the SVD factors, avoiding a second kernel bank.
}}

\subsection{Fractional-pixel oversampling via pre-shifted bases}
\label{sec:oversampling}

Visibility coordinates are sampled continuously in the $uv$-plane and therefore do not, in general, lie exactly on integer grid locations. For a continuous visibility coordinate $(u,v)$, each component can be decomposed into an integer grid index and a fractional offset,
\begin{equation}
u = \lfloor u \rfloor + \delta_u,
\qquad
v = \lfloor v \rfloor + \delta_v,
\end{equation}
where $\delta_u,\delta_v \in [0,1)$. These fractional offsets must be incorporated into the gridding convolution function to avoid decorrelation and interpolation error.

{\color{blue}
In conventional $w$-projection, oversampling is implemented by storing the full two-dimensional kernel at multiple sub-pixel offsets. An oversampling factor $N_{\mathrm{os}}$ in each dimension therefore increases the kernel-bank storage to $\mathcal{O}(N^2N_{\mathrm{os}}^2)$ per $w$-plane. {\color{blue}For example, \citet{Offringa2019Precision} report an oversampling factor of 4 for the \textsc{casa}/\textsc{tclean} $w$-projection gridder used in their study.}
}

{\color{blue}
In the SVD representation, oversampling is instead applied to the one-dimensional basis vectors. For each axis, the oversampling indices, fractional offsets, and shifted bases are defined together as
\begin{equation}
\begin{array}{@{}l@{}}
k_a \in \{0,\ldots,N_{\mathrm{os}}-1\},
\qquad
\delta_a = \dfrac{k_a}{N_{\mathrm{os}}},
\qquad
a\in\{u,v\},\\[3pt]
\widetilde{U}_w^{(k_u)}=\mathcal{S}_{\delta_u}[U_w],
\qquad
\widetilde{V}_w^{(k_v)}=\mathcal{S}_{\delta_v}[V_w].
\end{array}
\end{equation}

where $\mathcal{S}_{\delta}$ applies a one-dimensional fractional shift independently to each basis vector.
}

The fractional shift is implemented using the Fourier shift theorem. For a discrete vector $\mathbf{x} \in \mathbb{C}^N$,
\begin{equation}
\mathcal{S}_{\delta}[\mathbf{x}]
=
\mathcal{F}^{-1}
\left[
\mathcal{F}(\mathbf{x})
\odot
\exp\!\left(
-\frac{2\pi i k \delta}{N}
\right)
\right],
\end{equation}
where $k=0,1,\ldots,N-1$ indexes the discrete Fourier frequencies, $\mathcal{F}$ and $\mathcal{F}^{-1}$ denote the discrete Fourier transform and inverse transform, respectively, and $\odot$ denotes element-wise multiplication.

{\color{blue}
The oversampled kernel can therefore be reconstructed from shifted one-dimensional bases without storing every shifted two-dimensional kernel. For retained SVD rank $r_{\rm SVD}$, the respective storage requirements per $w$-plane scale as
\begin{equation}
\underbrace{\mathcal{O}(r_{\rm SVD}NN_{\mathrm{os}})}_{\text{SVD basis, per axis}}
\qquad\text{and}\qquad
\underbrace{\mathcal{O}(N^2N_{\mathrm{os}}^2)}_{\text{full 2D kernel bank}}.
\end{equation}
Since $r_{\rm SVD}\ll N$ in practice, this substantially reduces the oversampling overhead and permits larger oversampling factors.
}

During gridding, the fractional offsets $\delta_u$ and $\delta_v$ determine the oversampling-bin indices $k_u$ and $k_v$. The corresponding pre-shifted basis matrices are retrieved by direct lookup and combined to reconstruct the required kernel patch locally within the grid tile. This avoids any per-visibility FFTs or explicit shifting of the full two-dimensional kernel during imaging.

{\color{blue}
\subsection{Visibility Prediction Using the SVD-Compressed Kernel}
\label{sec:predict}

The predict operation, otherwise known as degridding, is the adjoint of gridding. It samples model visibilities from a model grid $G \in \mathbb{C}^{W \times W}$, which is the Fourier transform of the current model image. The predicted visibility is obtained by applying the forward measurement operator, corresponding to convolution with the complex-conjugate $w$-projection kernel,
\begin{equation}
V^{\mathrm{model}}_i =
\sum_{u,v} G(u,v)\, K_{w_i}^{*}(u,v),
\end{equation}
where $K_{w_i}^{*}$ denotes the complex conjugate of the corresponding $w$-projection kernel. Consistency between this forward operator and the adjoint gridding operator is required during iterative deconvolution.

BRISK applies the same pre-shifted SVD factors used during gridding, but in their conjugated form. Each grid tile evaluates only the part of the visibility footprint intersecting its local region, without constructing an intermediate two-dimensional kernel. When a footprint spans multiple tiles, the resulting tile-local contributions are accumulated into the same model visibility using atomic updates. Prediction therefore retains the reduced kernel storage and off-chip factor traffic of the gridding operation while preserving consistency between the forward and adjoint operators.
}

{\color{blue}
\subsection{Interpolation Between Adjacent $w$-planes}
\label{sec:wplane_interpolation}

In conventional $w$-projection, each visibility is commonly assigned to the nearest stored $w$-plane. This introduces a discretisation error because the true $w$ coordinate generally differs from the representative value of that plane. The error can be reduced by storing more planes, but for full two-dimensional kernels, the total kernel-bank memory grows as $\mathcal{O}(N_wN^2)$, where $N_w$ is the number of planes.

 The SVD-compressed representation instead permits efficient linear interpolation between the planes $w_b$ and $w_t$ that bracket the true coordinate $w_i$. Defining the lower-plane weight as $\alpha=(w_t-w_i)/(w_t-w_b)$, the interpolated kernel is
\begin{equation}
K_{w_i}(u,v)
\approx
\alpha K_{w_b}(u,v)
+
(1-\alpha)K_{w_t}(u,v).
\end{equation}

Rather than constructing this kernel explicitly, the contributions from the two planes are reconstructed separately from their compressed bases and weighted during gridding. This avoids interpolating the SVD bases, whose vectors may have inconsistent phases or signs between planes. 

Interpolation increases the arithmetic cost, since each visibility contributes to two neighbouring $w$-planes instead of one, but only the relevant one-dimensional basis-vector entries of the two planes are loaded. With full two-dimensional kernels, the same interpolation would require either two complete kernel patches per visibility or a bank of precomputed intermediate kernels, both of which add substantially to memory bandwidth and storage.

For negative $w$, interpolation is performed using the stored planes for $|w_i|$, followed by conjugation as described in Section~\ref{sec:kernel_recon}.

The same interpolation rule is also used during prediction. The model visibility is obtained by applying the adjoint operation to both neighbouring planes and combining their contributions using the same interpolation weights,
\begin{equation}
V^{\mathrm{model}}_i
\approx
\alpha V^{\mathrm{model}}_{i,b}
+
(1-\alpha)V^{\mathrm{model}}_{i,t},
\end{equation}
where $V^{\mathrm{model}}_{i,b}$ and $V^{\mathrm{model}}_{i,t}$ are the predicted visibility contributions obtained from the lower and upper $w$-planes, respectively.

Interpolation therefore removes the need to clamp visibilities to the nearest stored plane, and with it most of the error introduced by discrete $w$-plane sampling, without a prohibitively large number of stored planes; that this can be done at modest cost is a direct consequence of the compressed representation.
}

{\color{blue}
\section{BRISK imaging pipeline}
\label{sec:imager}

BRISK is an end-to-end CPU imaging pipeline written in \textsc{c++} and parallelised with \textsc{openmp}. It reads calibrated visibilities from \textsc{casa} Measurement Sets and writes the dirty image, point spread function (PSF), CLEAN model, residual image, and restored image as FITS products. Spectral channels are converted independently from metres to wavelengths and co-gridded at their natural $uv$ coordinates; the present implementation therefore provides narrow-band multi-frequency synthesis without per-channel spectral-index modelling.

For each non-negative $w$-plane, BRISK constructs the anti-aliased gridding convolution function (GCF), selects its support, and stores the truncated SVD representation described in Section~\ref{sec:SVDkernels}. Weighted visibilities are then gridded to form the dirty image and PSF. Deconvolution follows a Cotton--Schwab major/minor-cycle structure \citep{Schwab1984}: H\"ogbom CLEAN \citep{Hogbom1974} is used in the minor cycle, while each major cycle applies the SVD-based prediction operator of Section~\ref{sec:predict} to update the residual visibilities. Kernel construction, gridding and prediction, FFTs, and minor-cycle peak finding are multithreaded. The standard weighting, anti-aliasing, FFT, deconvolution, restoring-beam, and output details are given in Appendix~\ref{app:implementation}.

The two implementation choices that directly control the SVD kernel bank are retained here. First, the stored plane centres use a $\sqrt{|w|}$ distribution,
\begin{equation}
w_p = w_{\max}\left(\frac{p}{N_w-1}\right)^{\!2},
\qquad p=0,1,\ldots,N_w-1,
\end{equation}
where $w_{\max}$ is the smaller of the largest $|w|$ in the data and the geometric limit $1/(4|\Delta l|)$ set by the image sampling. Secondly, each GCF is cropped independently to the smallest radius $S$ for which
\begin{equation}
\frac{E_{\rm enclosed}(S)}{E_{\rm total}} \geq \kappa.
\end{equation}
The resulting full support width is $N=2S+1$. The support threshold $\kappa$ controls spatial truncation of the GCF, whereas the separate SVD threshold $\epsilon$ controls the number of retained singular modes. These choices allow support and rank to adapt independently across the $w$-plane bank; their accuracy-cost trade-offs are evaluated in Section~\ref{sec:param_validation}.
}

\section{Results and Analysis}
\label{sec:results}
 
\subsection{Datasets and Benchmarking Methodology}
\subsubsection{Measurement Sets}
\label{sec:measurement_sets}

Three primary measurement sets have been used to evaluate and benchmark our imaging pipeline against \textsc{wsclean} and \textsc{casa}. These datasets have been selected to span a range of complexity, from highly controlled simulations to a realistic sky model. All datasets used were produced with direction-dependent effects disabled \citep{Bhatnagar2008}, ensuring that imaging accuracy and error are attributed solely to the algorithm's ability to handle the $w$-term, rather than direction-varying antenna gains.

{\color{red}This is a controlled operator-level comparison: none of the pipelines compared here applies an $A$-term correction, so including a time- and direction-dependent primary beam would introduce an uncorrected error shared by all pipelines and would confound attribution of image differences to their handling of the $w$-term. The datasets therefore isolate the accuracy and cost of $w$-term gridding and prediction rather than testing a complete direction-dependent calibration and imaging chain.}

\textit{Dataset~1\,- OSKAR Point Source Grid:}
{\color{blue}
A regular $15 \times 15$ grid of 225 1\,Jy point sources simulated over an $8\degr$ FoV using \textsc{oskar} \citep{Mort2010} with an SKA1-Mid telescope model comprising 63 stations (Figure~\ref{fig:15by15grid}). The uniform flux, known source positions, and maximum $w$-value of $35502.8\lambda$ make this an ideal testbed for accuracy metric tests in the wide-field imaging regime: any $w$-correction, gridding, or imaging errors become immediately visible in the recovered image. Simulation parameters are summarised in Table~\ref{tab:oskar_params}.
}
 
\textit{Dataset~2\,- OSKAR Random Point Sources:}
50 randomly positioned point sources, simulated under the same \textsc{oskar} settings and telescope model as Dataset~1 but with a more realistic observational configuration: 50 frequency channels spanning a 100\,MHz bandwidth, 800 time steps and a 2\,s time smearing averaging per dump. This random source distribution makes deconvolution more challenging than the grid in Dataset~1, and the increased spectral and temporal sampling provides a measurement set that is more representative of those collected by real radio telescope arrays. Simulation parameters are summarised in Table~\ref{tab:oskar_params}.
 
For both OSKAR datasets, imaging accuracy and runtime are fixed at a constant $8\degr$ FoV. Doubling the image size, therefore, halves the pixel scale, allowing imaging accuracy and speed to be assessed as a function of image resolution alone.
 
\textit{Dataset~3\, - SKAO GLEAM Simulation:}
A set of three measurement sets (small, medium, large) drawn from a publicly available SKAO Simulations repository\footnote{\url{https://gitlab.com/ska-telescope/sim/sim-datasets}}, based on a sky model derived from the GLEAM catalogue over the EoR1 target field. The three configurations use the full SKA1-Low array of 512 stations, giving 130,816 instantaneous baselines per dump, and maintain a constant angular resolution of 2\,arcsec\,pixel$^{-1}$ while varying in field of view and image size. Because Dataset~3 combines a real catalogue, the full SKA1-Low array, and observation lengths drawn from typical SKA survey strategies, it provides the most realistic test of BRISK performance under conditions representative of next-generation surveys. Simulation parameters are summarised in Table~\ref{tab:gleam_params}. {\color{blue}Only the $XX$ polarisation is provided, so \textsc{idg}, which requires all four, is omitted for all GLEAM runs.}
 
\subsubsection{Benchmarking Metrics}
\label{sec:benchmarking_metrics}
 
{\color{blue}
To assess imaging performance, we compare both the reconstruction fidelity and computational efficiency of our pipeline against \textsc{casa}'s \textsc{tclean} and three gridders available in \textsc{wsclean}: \textsc{wstacking}, \textsc{wgridder}, and \textsc{image domain gridding (idg)}. Each \textsc{wsclean} gridder is run in two configurations: the standard CLEAN pipeline and the \textsc{parallel deconvolution} (PD) mode. Because BRISK is targeted at next-generation arrays where image accuracy, runtime and memory are all constraints, image-quality metrics are combined with measurements of wall-clock runtime and peak memory usage. No single image-quality metric is sufficient on its own, so three complementary measures are used: the mean source-flux standard error, the residual-image noise floor, and the ITW-SSIM \citep{LiArmour2022} score of the final model image.
}
 
{\color{blue}
\textit{Mean source-flux standard error} is the mean recovered peak flux density over all detected sources, with the standard deviation across sources reported as a measure of how uniformly each imager recovers brightness across the field. Sources are extracted from the restored images with SoFiA~2 \citep{Westmeier2021}. For the \textsc{oskar} simulations, the true fluxes are known exactly from the input sky model; for the GLEAM datasets, the detections are matched against the input catalogue to count the sources recovered.

\textit{Residual image noise floor} is the rms of the residual image after deconvolution. Lower values indicate a more complete deconvolution; any remaining structure reflects imaging artefacts or incomplete source subtraction.

\textit{Intensity-weighted structural similarity} (ITW-SSIM, \citealt{LiArmour2022}) measures the structural agreement between two images on a scale of 0 to 1, weighting the standard SSIM by local intensity so that pixels containing emission dominate. It is used to quantify how closely each pipeline reproduces the sky structure recovered by the others: the values in Tables~\ref{tab:gridder_accuracy} and~\ref{tab:gleam_accuracy} are computed against the non-interpolated BRISK model image, which is why that row carries no value. The comparison is made on the CLEAN model images rather than the restored or residual images, which isolates gridding and deconvolution accuracy from the small differences in fitted restoring beam between imagers.

\textit{Wall-clock runtime} and \textit{peak memory usage} are recorded for every run to assess scalability. Runtime covers the entire run and, where relevant, is broken down by algorithmic stage; peak memory is the maximum memory footprint recorded during the run. Both are reported as functions of image size, which is the axis along which SKA-scale imaging becomes expensive.
}
 
\subsubsection{Hardware and Software Configuration}
 
{\color{blue}
All benchmarking results were obtained on a single dual-socket AMD Genoa compute node with two AMD EPYC 9634 processors at 2.25\,GHz (168 cores in total) and 2.2\,TiB of system memory. \textsc{wsclean} version 3.5 and \textsc{casa} version 6.5 were used for comparison, each with 48 cores. For BRISK, 48 cores were used for gridding and deconvolution, and 16 for the FFTW Fourier transforms, which we found to be the optimal FFTW thread count. All \textsc{wgridder} runs used the default accuracy setting of $10^{-4}$, which permits a single-precision implementation; the stricter $10^{-5}$ setting forces double-precision arithmetic and increases both peak memory and wall time by approximately 50\%, so the results shown represent \textsc{wgridder} in its most efficient configuration. BRISK was run with the support threshold $\kappa = 99.9\%$ selected in Section~\ref{sec:param_validation} and the rank threshold $\epsilon = 0.99999 $  for every benchmark. All reported wall times are cold-start, end-to-end measurements for every pipeline, from reading the measurement set to writing the restored image: they include the construction of the $w$-kernels (and, for BRISK, their SVD compression) in every run, and no kernel cache was used by any code. {\color{red}A stage-by-stage runtime breakdown is given in Section~\ref{sec:gleam_runtime}.}
}

The subsequent sections also refer to the parallel deconvolution (PD) setting in \textsc{wsclean}, with the sub-image size set to 1024 pixels for all tests. This divides the full dirty image into a set of sub-images, with the splitting boundaries chosen using a Dijkstra-based procedure that minimises the absolute path through the image \citep{OffringaPDDocs}. The resulting sub-images can then be deconvolved independently on separate CPU cores, reducing the cost of the minor-cycle subtraction step. This is particularly beneficial when CLEAN dominates the total runtime, as occurs for large images with many active components.  However, as shown in Section~\ref{sec:oskar_accuracy}, this acceleration is not always accuracy-preserving at the largest image sizes.
 
\subsection{Parameter Validation for BRISK}
\label{sec:param_validation}
 
\begin{figure*}
    \centering
    \subcaptionbox{\label{fig:ssim_energykeep}}{\includegraphics[height=5.2cm]{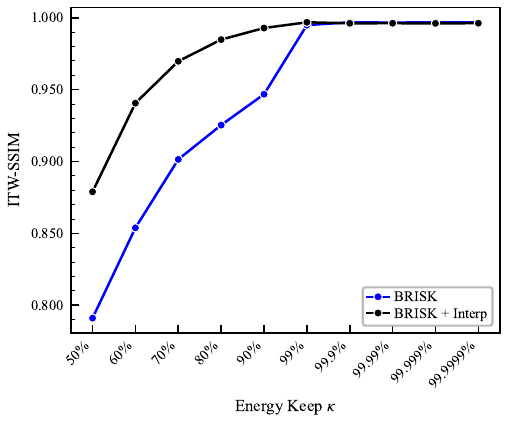}}\hspace{4pt}%
    \subcaptionbox{\label{fig:timing_energykeep}}{\includegraphics[height=5.2cm]{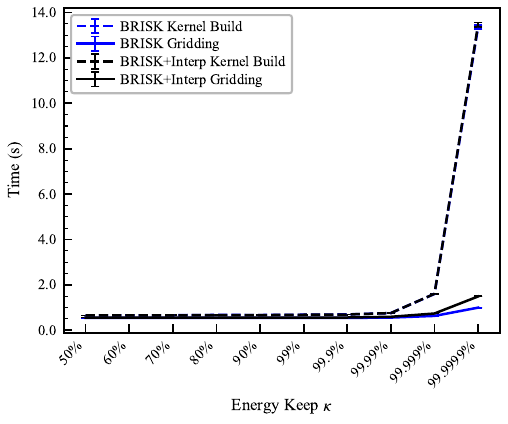}}\\[4pt]
    \subcaptionbox{\label{fig:ssim_wprojplanes}}{\includegraphics[height=5.2cm]{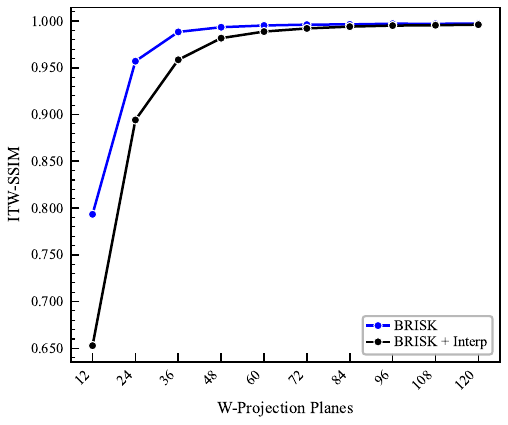}}\hspace{4pt}%
    \subcaptionbox{\label{fig:timing_wprojplanes}}{\includegraphics[height=5.2cm]{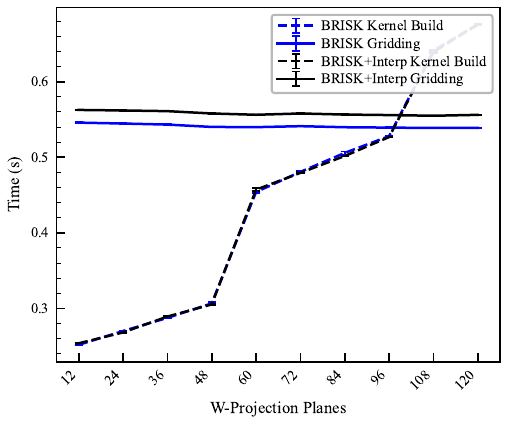}}
    \caption{Accuracy-cost trade-off for the energy-retention threshold $\kappa$ and the number of stored $w$-planes $N_w$, measured against a high-fidelity \textsc{wgridder} reference image. \textit{(a)} ITW-SSIM versus $\kappa$. \textit{(b)} Kernel-build and gridding times versus $\kappa$. \textit{(c)} ITW-SSIM versus $N_w$; note that interpolation underperforms nearest-plane at very low $N_w$. \textit{(d)} Kernel-build and gridding times versus $N_w$.}
    \label{fig:accuracy_cost}
\end{figure*}

Figure~\ref{fig:accuracy_cost} shows the accuracy-cost trade-off for two key tuning parameters in our imager: the energy-retention threshold $\kappa$ used during kernel construction, and the number of $w$-projection planes, $N_w$. To measure image fidelity, we use the ITW-SSIM benchmarking metric with a high-accuracy \textsc{wsclean}/\textsc{wgridder} dirty image of Dataset~1 as the reference image.  Panels~\ref{fig:ssim_energykeep} and~\ref{fig:ssim_wprojplanes} show the dependence of ITW-SSIM on $\kappa$ and $N_w$, respectively, both with and without the adjacent-plane interpolation scheme. Panels~\ref{fig:timing_energykeep} and~\ref{fig:timing_wprojplanes} show the corresponding kernel-build and grid-to-image times.

Panel~\ref{fig:ssim_energykeep} shows that increasing the retained kernel energy rapidly improves image fidelity, with the curve beginning to saturate by $\kappa=99.9\%$. Beyond this point, retaining additional kernel energy adds very little extra information to the dirty image. The interpolated case performs noticeably better at lower values of $\kappa$, where the GCF representation is more approximate and nearest-plane assignment introduces an additional discretisation error in the $w$ coordinate. By splitting each visibility between the two adjacent stored $w$-planes, interpolation reduces this discontinuous $w$-quantisation error and gives a dirty image that agrees more closely with the \textsc{wgridder} reference. Once the kernel representation itself is sufficiently accurate and the two curves converge, the remaining difference caused by the interpolation choice becomes small for this dataset.

{\color{blue}
Panel~\ref{fig:ssim_wprojplanes} shows the dependence on the number of stored $w$-planes. Without interpolation, increasing $N_w$ from 12 to 84 planes produces a strong gain in ITW-SSIM, after which little further improvement is obtained. This is the expected behaviour of nearest-plane assignment: a visibility whose $w$ differs from that of its assigned plane by $\delta w$ carries an uncorrected phase error of $2\pi\,\delta w\,(\sqrt{1-l^2-m^2}-1)$, and for a uniformly sampled $w$ grid, with $\delta w \la w_{\max}/[2(N_w-1)]$, the maximum error at the edge of the field scales as
\begin{equation}
\epsilon_\phi
\sim
\frac{\pi\,w_{\max}\,\theta_{\max}^2}
     {2\,(N_w-1)},
\label{eq:nw_min}
\end{equation}
where $\theta_{\max}$ is the half-width of the field of view in radians \citep{Cornwell2008, Offringa2014}. The inverse dependence on $N_w$ explains the rapid initial improvement, followed by saturation once the residual phase error falls below the other image-error terms.
}

Although adjacent-plane interpolation is designed to reduce the continuous-$w$ approximation error, it is not guaranteed to improve the image when the $w$-plane sampling is very coarse. In this regime, neighbouring stored kernels can be separated by a large interval in $w$, and the GCF does not vary linearly between them. The interpolated kernel is then a weighted average of two valid kernels, but is not necessarily a better approximation to the true intermediate-$w$ kernel than selecting the nearest stored plane. This effect is amplified by our adopted $\sqrt{|w|}$ plane distribution,
\begin{equation}
w_p = w_{\max}\left(\frac{p}{N_w-1}\right)^2 ,
\end{equation}
for which the spacing between adjacent planes increases with $p$:
\begin{equation}
\Delta w_p
=
w_{p+1}-w_p
=
w_{\max}\frac{2p+1}{(N_w-1)^2}.
\end{equation}
Thus, for small $N_w$, the high-$w$ end of the kernel bank is sampled by widely separated planes, precisely where the kernels are more oscillatory and less well described by a linear blend. This explains why interpolation underperforms nearest-plane assignment at very low plane counts, while the two approaches converge once the $w$-plane grid is sufficiently dense. The result indicates that interpolation is most useful as a refinement to an adequately sampled $w$-plane grid, rather than as a substitute for using a sufficient number of stored planes.  Interpolation therefore reduces the residual error caused by discretising the continuous $w$ coordinate, producing a gridding operator that more closely approximates the true measurement equation, improving the accuracy of the dirty image and PSF, and the forward prediction step used during Cotton-Schwab major cycles. As a result, the residual visibilities after each major cycle contain less unmodelled $w$-term error, allowing the deconvolution to proceed to a deeper and more accurate CLEAN model before artefacts from the approximate measurement operator become limiting. The impact of this improved operator accuracy on the final restored image is demonstrated experimentally in Section~\ref{sec:oskar_accuracy}.

The corresponding timing figures in panels~\ref{fig:timing_energykeep} and~\ref{fig:timing_wprojplanes} show how the kernel-build time and the combined grid-to-image time vary with $\kappa$ and $N_w$. For both sweeps, increasing $\kappa$ or increasing $N_w$ primarily affects the kernel-construction stage. Increasing $\kappa$ requires retaining wider or more accurate GCF support during kernel construction, while increasing $N_w$ requires constructing and compressing more $w$-planes. The grid-to-image time is comparatively flat because the gridding stage operates on the compressed SVD basis and reconstructs only the required local kernel patches on the fly. When adjacent-plane interpolation is enabled, the gridding and prediction cost increase because each visibility contributes to two neighbouring $w$-planes instead of one. However, this additional cost is far smaller than it would be for full two-dimensional kernels, as mentioned in Section \ref{sec:wplane_interpolation}.

{\color{blue}
Together, the trends in Figure~\ref{fig:accuracy_cost} suggest an operating point of $\kappa = 99.9\%$ for the subsequent benchmarks: this retains essentially all of the achievable dirty-image accuracy while avoiding the kernel-construction cost of larger supports. Interpolation is treated throughout as an accuracy-enhancing option applied on top of an adequately sampled $w$-plane grid.
}

The number of $w$-planes is set by an automatic recommender within the pipeline, which uses the relation
\begin{equation}
N_w =
\left\lceil
2\sigma w_{\max}
\left(
1 - \sqrt{1 - 2l_{\max}^2}
\right)
\right\rceil ,
\label{eq:nw_recommend}
\end{equation}
where $\sigma = \pi/2$ is a kernel-aware oversampling constant, half the strict phase-budget of \citet{Offringa2014} since the convolutional $w$-kernel absorbs sub-plane phase; $l_{\max} = W\Delta l/2$ is the half-field of view in direction cosines; and
\begin{equation}
w_{\max}
=
\min\left(\max |w_{\rm data}|,\frac{0.25}{|\Delta l|}\right)
\end{equation}
{\color{blue}is the effective $w$ range adopted by BRISK; the $0.25/|\Delta l|$ ceiling is an implementation-specific pixel-scale cap. The dependence of $N_w$ on $w$ range and field extent follows the phase-error argument of \citet{Offringa2014}, which that study also used to set matched $w$-layer and $w$-projection-plane counts for \textsc{wsclean} and \textsc{casa}.} 

For the OSKAR scaling benchmarks in this work, we found that this recommended number of $w$-layers is conservative: empirical convergence of ITW-SSIM occurs at a lower $N_w$ than recommended. We therefore use $N_w = 128$ for $W \leq 8192$, $N_w = 256$ at $W=16384$, $N_w = 512$ at $W=32768$, and $N_w = 800$ at $W=65536$ and larger. The same $N_w$ schedule is used for the \textsc{casa} $w$-projection comparisons up to $W=8192$ and for \textsc{wsclean}'s \textsc{wstacking} comparisons up to $W=32768$, ensuring that all imagers operating on the same $w$-correction principle receive an equivalent number of planes. Both comparison pipelines became too slow to run at larger image sizes than mentioned. The \textsc{wgridder} and \textsc{idg} pipelines determine their own internal $w$-handling parameters automatically and are run with their default settings.
 
\subsection{Performance on Simulated OSKAR Data}
\label{sec:oskar_results}

\begin{figure*}[!t]
    \centering
    \includegraphics[width=0.97\textwidth]{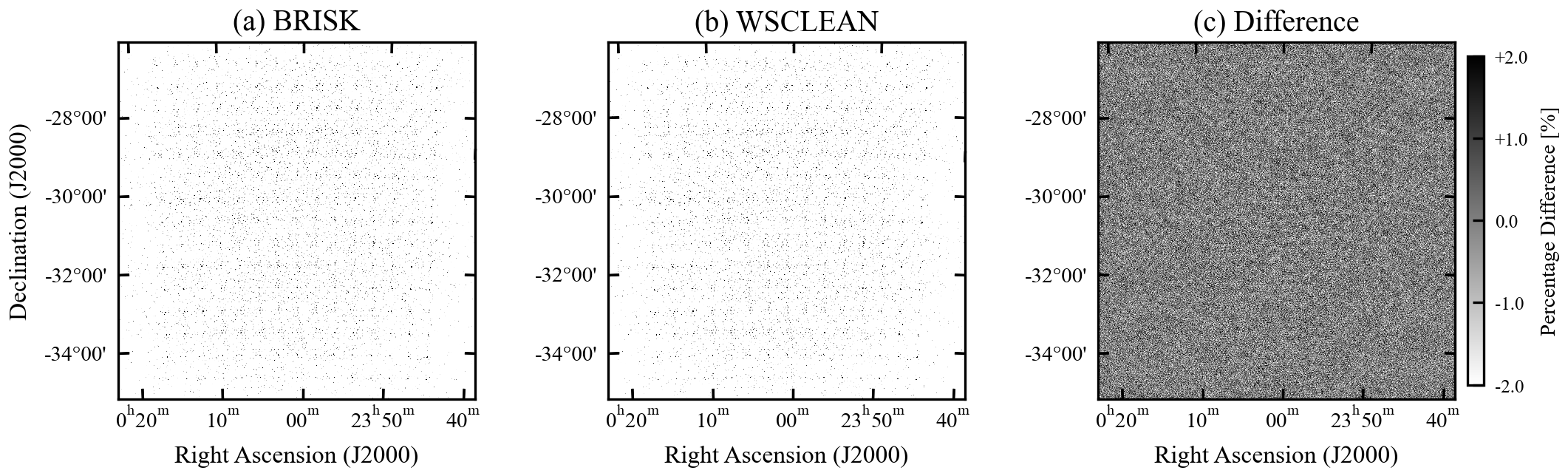}
    \caption{Dirty images of Dataset~1 (225 unit-flux point sources on a $15\times15$ grid over an $8\degr$ FoV) from (a) BRISK with interpolation and (b) \textsc{wsclean}'s \textsc{wgridder}, with (c) their difference in per cent of the dirty-image peak.}
    \label{fig:15by15grid}
\end{figure*}

\subsubsection{Reconstruction Accuracy}
\label{sec:oskar_accuracy}
Table~\ref{tab:gridder_accuracy} summarises the source-extraction and image-fidelity statistics for all pipelines as a function of image size, using the metrics of Section~\ref{sec:benchmarking_metrics}. BRISK results are shown both with and without interpolation to isolate the effect of the interpolation stage. {\color{blue}Figure~\ref{fig:15by15grid} shows the agreement directly in the image domain: the BRISK and \textsc{wsclean} dirty images of Dataset~1 differ by an rms of $0.84\%$ of the dirty-image peak across the field, with no structure at the positions of the sources.}
 
At the smallest image size, $1024^2$ pixels, all methods exhibit noticeable peak-flux bias: the BRISK variants slightly overestimate the peak, \textsc{wstacking}, \textsc{wgridder}, and \textsc{idg} underestimate it by $18$--$21\%$, and \textsc{casa} performs best at {\color{blue}$+2.43\%$}. This is expected, as at this resolution the image is severely under-sampled relative to the SKA1-Mid AA2 synthesised beam, so the recovered peak is dominated by each imager's sub-pixel and gridding/degridding interpolation rather than by the asymptotic accuracy of the $w$-correction. ITW-SSIM nonetheless remains close to unity for all methods, indicating that the global source distribution is recovered well.
 
{\color{blue}At $2048^2$ and $4096^2$ pixels the pixel scale becomes better matched to the synthesised beam, peak-flux biases fall to the few-percent level, and most methods retain ITW-SSIM above 0.99; the exception is \textsc{casa} at $4096^2$, with 0.978. At $4096^2$, interpolation reduces the BRISK mean peak-flux bias from $-1.28\%$ to $+0.10\%$, while the source-to-source scatter changes only slightly, from 0.0062 to 0.0078 Jy/beam.}
 
Beyond $4096^2$ pixels the methods separate. The \textsc{wstacking}-based methods degrade progressively, reaching a $\sim-17\%$ peak-flux bias by $32768^2$ pixels with ITW-SSIM falling to 0.923, as a fixed $N_w$ schedule no longer provides sufficient $w$-correction across the field at finer pixel scales. \textsc{wgridder} and \textsc{idg} remain accurate in their standard modes, holding sub-1\% bias and ITW-SSIM $\approx0.996$ at $32768^2$ pixels. The interpolated BRISK follows this same high-accuracy trend: once the image is sufficiently sampled, interpolation removes the systematic flux loss seen in the non-interpolated results, and at $32768^2$ pixels it reaches a peak-flux bias of only {\color{blue}$-0.06\%$} (against $-0.55\%$ for \textsc{wgridder} and $-0.57\%$ for \textsc{idg}) with ITW-SSIM of 0.997. With interpolation enabled, the BRISK reconstruction is therefore not merely stable with image size but competitive with the most accurate production gridders tested.

{\color{blue}
The PD variants of \textsc{wgridder} and \textsc{idg} improve runtime (Section~\ref{sec:oskar_runtime}) but become less reliable at the largest image sizes: the effect is visible at $16384^2$ pixels and severe at $32768^2$, where PD inflates the source-to-source scatter by an order of magnitude and drops ITW-SSIM to 0.891 (\textsc{wgridder}) and 0.856 (\textsc{idg}). The contribution of stopping depth is constrained by the equal-noise-floor experiment in Section~\ref{sec:oskar_runtime}.
}

For image sizes of $65536^2$ pixels and above, \textit{only} BRISK provided a practical route to completion. The standard gridders in \textsc{wsclean} exceeded our 24-hour wall-clock budget, and for the PD variants, we observed divergence during the first major cycle, with the peak residual growing as parallel minor-cycle subtractions accumulated negative PSF sidelobe structure faster than later iterations could cancel it. 

{\color{blue}
Repeating the test on the random source distribution of Dataset~2 confirmed that the divergence is not specific to the structured grid: the deconvolution progressed through 10 major cycles before the peak residual began to grow, but it again failed to reach the termination threshold and produced no final image. Under the settings tested, this points to a stability limitation of independent sub-image minor cycles in very large wide-field images, where extended PSF support and many active components combine to give inconsistent subtraction across the image.
}
 
{\color{blue}BRISK is stable across the full range to $131072^2$ pixels even without interpolation, with the recovered peak flux staying within a few percent of unity and the scatter remaining below 0.0066 Jy/beam from $32768^2$ pixels upwards. With interpolation, the large-image accuracy improves further, reaching a scatter of 0.0038 Jy/beam and a noise floor of 0.52 mJy/beam at $131072^2$ pixels (Table~\ref{tab:gridder_accuracy}). Together with the runtime and memory results below, this establishes BRISK as the most robust combination of accuracy, stability, and scalability in the wide-image regime.}
 
\subsubsection{Runtime Scaling}
\label{sec:oskar_runtime}
\begin{figure}[!b]
    \centering
    \includegraphics[width=\linewidth]{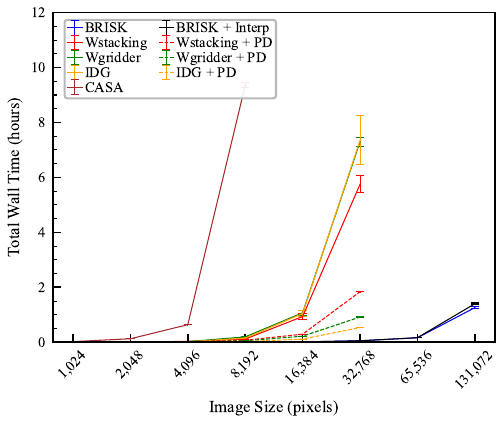}
    \caption{Total wall time versus image size at a fixed $8\degr$ FoV (Dataset~1), with each pipeline run to its own auto-threshold. Solid lines are standard CLEAN; dashed lines are \textsc{wsclean} with parallel deconvolution. Error bars show run-to-run variation.}
    \label{fig:runtime_comparison}
\end{figure}

Figure~\ref{fig:runtime_comparison} shows the wall-time scaling of the gridders as a function of image size at a fixed $8\degr$ field of view, with each gridder run to its own automatically estimated noise floor.

BRISK scales far more shallowly than the comparison methods, and the separation grows rapidly with image size. At $131072^2$ pixels it produces a cleaned image in under {\color{blue} 90 minutes} on a single dual-socket AMD node, whereas none of the comparison imagers could be pushed beyond $32768^2$ pixels without either failing to converge or exceeding the 24-hour wall-time budget.
\begin{table*}[!t]
\color{blue}
\centering
\caption{\color{blue}Overall wall-time speed-up factor of BRISK relative to the interpolated BRISK and other standard and parallel-deconvolution gridders across scaled image sizes. Values larger than unity indicate that the corresponding comparison method is slower than BRISK. Uncertainties are quoted as $1\sigma$ propagated errors.}
\label{tab:runtime_speedup}

\setlength{\tabcolsep}{4pt}

\resizebox{\textwidth}{!}{%
\begin{tabular}{lcccccccc}
\hline\hline
Image Size &
BRISK + Interp. &
WSTACKING &
WSTACKING (PD) &
WGRIDDER &
WGRIDDER (PD) &
IDG &
IDG (PD) &
CASA \\
\hline

1024  &
$1.08 \pm 0.14$ &
$9.01 \pm 1.67$ &
$6.80 \pm 2.40$ &
$3.34 \pm 1.13$ &
$2.59 \pm 0.19$ &
$11.33 \pm 0.85$ &
$11.84 \pm 1.52$ &
$58.56 \pm 4.41$ \\

2048  &
$0.94 \pm 0.11$ &
$11.13 \pm 1.19$ &
$9.22 \pm 1.02$ &
$12.31 \pm 1.83$ &
$3.44 \pm 0.59$ &
$10.96 \pm 1.21$ &
$11.70 \pm 1.87$ &
$167.12 \pm 16.81$ \\

4096  &
$1.26 \pm 0.09$ &
$27.16 \pm 2.42$ &
$16.72 \pm 1.13$ &
$19.22 \pm 1.63$ &
$12.35 \pm 1.41$ &
$14.59 \pm 1.57$ &
$10.28 \pm 0.90$ &
$517.38 \pm 35.25$ \\

8192  &
$2.13 \pm 0.11$ &
$30.61 \pm 0.74$ &
$18.82 \pm 0.45$ &
$51.48 \pm 5.04$ &
$14.41 \pm 0.46$ &
$38.82 \pm 4.81$ &
$6.73 \pm 0.24$ &
$2443.80 \pm 60.90$ \\

16384 &
$1.42 \pm 0.16$ &
$81.65 \pm 12.77$ &
$25.73 \pm 2.38$ &
$93.52 \pm 11.63$ &
$19.39 \pm 1.81$ &
$91.32 \pm 11.79$ &
$10.23 \pm 0.95$ &
-- \\

32768 &
$0.96 \pm 0.03$ &
$103.01 \pm 6.25$ &
$32.95 \pm 0.98$ &
$130.71 \pm 4.75$ &
$16.46 \pm 0.49$ &
$131.64 \pm 16.50$ &
$9.60 \pm 0.30$ &
-- \\

\hline
\hline
\end{tabular}%
}
\end{table*}

{\color{blue}
The corresponding speed-up factors are summarised in Table~\ref{tab:runtime_speedup}. Against \textsc{wsclean}'s standard gridders the advantage is large at every image size and grows with it, reaching $103.0\times$ over \textsc{wstacking}, $130.7\times$ over \textsc{wgridder}, and $131.6\times$ over \textsc{idg} at $32768^2$ pixels. \textsc{casa} is slower still: at $8192^2$ pixels, the largest size at which it completed, BRISK is $\sim2444\times$ faster, and beyond this \textsc{casa} becomes impractical.
}
 
\textsc{idg}'s comparatively poor showing at the smallest image sizes is an amortisation effect. Rather than precomputing oversampled convolution functions, \textsc{idg} groups visibilities into subgrids, forms and Fourier-transforms low-resolution images, and accumulates them onto the full $uv$ grid \citep{vanderTol2018}, incurring fixed batching and orchestration costs. These dominate when the image-dependent work is small and are amortised as the image grows, which is why \textsc{idg} becomes more competitive at larger sizes.
 
Interpolation increases the cost of each gridding operation but does not consistently lengthen the end-to-end runtime: across the image-size range the {\color{blue}interpolated BRISK runtime ranges from 6\% faster to a factor of 2.1 slower than the non-interpolated BRISK}. This variation is driven by the interaction between interpolation accuracy and the CLEAN stopping criteria rather than the interpolation overhead itself — by lowering the estimated residual noise floor, interpolation can let CLEAN proceed into an additional major cycle for some image sizes while reaching the threshold faster for others. The overhead is therefore negligible relative to the standard \textsc{wsclean} gridders, while the accuracy gains of Section~\ref{sec:oskar_accuracy} remain.
 
{\color{blue}
The advantage persists against \textsc{wsclean}'s PD mode. The factors are naturally smaller, since PD reduces the comparison methods' deconvolution cost, but BRISK remains $33.0\times$, $16.5\times$, and $9.6\times$ faster than \textsc{wstacking} (PD), \textsc{wgridder} (PD), and \textsc{idg} (PD) at $32768^2$ pixels, and these PD runtimes come with the accuracy degradation noted in Section~\ref{sec:oskar_accuracy}.
}

\begin{table*}[!b]
\centering
\captionsetup{font=small,skip=3pt}
\caption{{\color{blue}Runtime and source-extraction accuracy as a function of image size when all methods are run to the stopping level defined by the non-interpolated BRISK result. Peak-flux entries give the mean recovered source peak in Jy/beam, with the minimum and maximum recovered values in square brackets. BRISK runtimes are application runtimes averaged over successful repeats where available; the comparison methods are single-run wall times. PD denotes the \textsc{wsclean} parallel-deconvolution mode.}}
\label{tab:equal_noisefloor_runtime}
\footnotesize
\setlength{\tabcolsep}{3pt}
\begin{tabular}{l c c c c}
\hline\hline
 & BRISK & BRISK + Interp. & WGRIDDER (PD) & IDG (PD) \\
\hline
\multicolumn{5}{l}{\textit{\textbf{Image Size 4096}}} \\
\hline
{\color{blue}Runtime (s)} & {\color{blue}6.92} & {\color{blue}5.03} & {\color{blue}46.3} & {\color{blue}56.6}\\
Mean Peak Flux (Jy/b) & {\color{blue}0.9872 [0.9682, 0.9996]} & {\color{blue}1.0062 [0.9857, 1.0284]} & {\color{blue}0.9963 [0.9432, 1.0041]} & {\color{blue}0.9963 [0.9418, 1.0040]} \\
Std. Deviation (Jy/b) & {\color{blue}0.0062} & {\color{blue}0.0063} & {\color{blue}0.0098} & {\color{blue}0.0099} \\
\hline
\multicolumn{5}{l}{\textit{\textbf{Image Size 8192}}} \\
\hline
{\color{blue}Runtime (s)} & {\color{blue}15.83} & {\color{blue}8.14} & {\color{blue}92.9} & {\color{blue}62.4} \\
Mean Peak Flux (Jy/b) & {\color{blue}0.9730 [0.9358, 0.9874]} & {\color{blue}1.0121 [0.9980, 1.0571]} & {\color{blue}0.9988 [0.9959, 1.0018]} & {\color{blue}0.9988 [0.9960, 1.0017]} \\
Std. Deviation (Jy/b) & {\color{blue}0.0096} & {\color{blue}0.0122} & {\color{blue}0.0010} & {\color{blue}0.0010} \\
\hline
\multicolumn{5}{l}{\textit{\textbf{Image Size 16384}}} \\
\hline
{\color{blue}Runtime (s)} & {\color{blue}49.61} & {\color{blue}33.91} & {\color{blue}487.5} & {\color{blue}239.8} \\
Mean Peak Flux (Jy/b) & {\color{blue}0.9948 [0.9735, 1.0090]} & {\color{blue}1.0056 [0.9738, 1.0628]} & {\color{blue}0.9994 [0.9874, 1.0172]} & {\color{blue}0.9996 [0.9881, 1.0174]} \\
Std. Deviation (Jy/b) & {\color{blue}0.0065} & {\color{blue}0.0164} & {\color{blue}0.0053} & {\color{blue}0.0053} \\
\hline
\multicolumn{5}{l}{\textit{\textbf{Image Size 32768}}} \\
\hline
{\color{blue}Runtime (s)} & {\color{blue}170.52} & {\color{blue}131.29} & {\color{blue}2359.5} & {\color{blue}682.3} \\
Mean Peak Flux (Jy/b) & {\color{blue}0.9918 [0.9729, 1.0098]} & {\color{blue}1.0034 [0.9866, 1.0189]} & {\color{blue}1.0068 [0.9921, 1.0172]} & {\color{blue}1.0072 [0.9929, 1.0150]} \\
Std. Deviation (Jy/b) & {\color{blue}0.0066} & {\color{blue}0.0064} & {\color{blue}0.0040} & {\color{blue}0.0039} \\
\hline\hline
\end{tabular}
\end{table*}

{\color{blue}Table~\ref{tab:equal_noisefloor_runtime} compares the fastest pipelines at a stopping level defined by the non-interpolated BRISK result. The interpolated BRISK remains substantially faster than both PD variants at every image size, completing in 131.29\,s at $32768^2$ pixels against 682.3\,s for \textsc{idg} (PD) and 2359.5\,s for \textsc{wgridder} (PD), speed-ups of $5.2\times$ and $18.0\times$, respectively. The wall-time advantage therefore persists when the stopping depths are matched.}
 
{\color{blue}The matched-depth results also constrain the origin of the PD accuracy degradation. The PD means remain within 0.72\% of the true 1-Jy peak at every tested size, and, most importantly, at $32768^2$ the matched-depth scatter is only 0.0040 Jy/beam for \textsc{wgridder} (PD) and 0.0039 Jy/beam for \textsc{idg} (PD), compared with 0.0456 and 0.0858 Jy/beam in their deeper auto-threshold runs in Table~\ref{tab:gridder_accuracy}. This strongly associates the large-image degradation with the depth of parallel deconvolution, and is consistent with sidelobe-induced cross-talk between independent facet minor cycles. Because this experiment changes stopping depth rather than directly measuring cross-facet subtraction errors, however, it supports rather than proves that mechanism.}
 
\begin{figure}
    \centering
    \includegraphics[width=\linewidth]{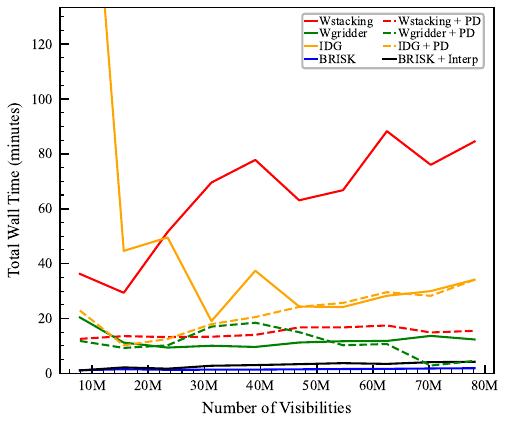}
    \caption{Wall time versus visibility count at a fixed $8192^2$ image size, obtained by imaging increasing frequency-band subsets of Dataset~2. The BRISK curves are essentially flat, indicating that runtime is image-size- rather than visibility-limited in this regime. The first \textsc{idg} point ($\sim237$\,min) is truncated by the upper axis limit.}
    \label{fig:vis_walltime}
\end{figure}

Figure~\ref{fig:vis_walltime} shows wall time as a function of visibility count, at a fixed $8192^2$ image size with a 4-arcsec pixel scale, obtained by imaging increasing frequency-band subsets of Dataset~2 spanning $\sim$8-78 million visibilities (\textsc{casa} is excluded for its prohibitive runtime). BRISK holds an essentially flat wall time of $\sim$70--110\,s across the full range — an order of magnitude below the next-fastest pipelines, \textsc{wgridder} and \textsc{wgridder} (PD) at $\sim$500--800\,s. This weak dependence on visibility count shows that the dominant costs here are not per-visibility but are set by image-size-dependent operations (the FFTs) and the compact SVD-kernel reconstruction. By contrast, \textsc{wstacking} increases more clearly with visibility count and \textsc{idg} shows substantial non-monotonic scatter, indicating sensitivity to the visibility distribution itself.
 
Together, Figures~\ref{fig:runtime_comparison} and~\ref{fig:vis_walltime} show that the BRISK's wall-time advantage holds along both principal scaling axes — image size and visibility count — by reducing the cost of applying wide-field convolution functions while avoiding the memory and deconvolution bottlenecks that limit the comparison methods. The advantage grows with image size and persists at a fixed noise floor.

{\color{red}
 \subsubsection{Runtime Comparisons Against \textsc{idg} GPU and Hybrid}
Table~\ref{tab:idg_gpu_breakdown} compares the stage-level runtimes at $16384^2$ pixels. All runs used 48 cores of an Intel Xeon Gold 6548N processor, while the GPU and hybrid modes of \textsc{idg} additionally used one 48-GB NVIDIA L40S. GPU acceleration reduced inversion by $3.7\times$ and prediction by $2.6\times$ relative to \textsc{idg} CPU, but CPU-side deconvolution still occupied $82.6$--$82.9\%$ of wall time, leaving both modes approximately $8.7\times$ slower than BRISK.
 }
\subsubsection{Peak Memory Usage}
\label{sec:oskar_memory}

\begin{figure}[!htbp]
\color{blue}
\centering
\includegraphics[width=0.82\linewidth]{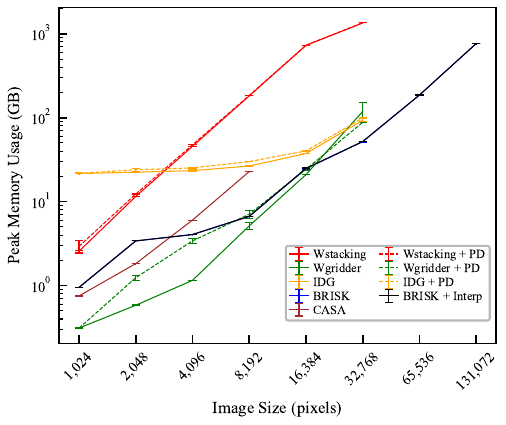}
\caption{\color{blue}Peak memory usage as a function of image size for all imaging pipelines, on Dataset~1 with an $8\degr$ field of view. Dashed curves indicate \textsc{wsclean} runs with parallel deconvolution (PD) enabled; solid curves are standard CLEAN. Error bars span the spread of repeated runs.}
\label{fig:runtime_memory}
\end{figure}

Figure~\ref{fig:runtime_memory} shows peak memory against image size; all methods grow with image size, but the scaling differs sharply. \textsc{wstacking} is the most demanding, storing multiple image-sized $w$-layers and exceeding 1\,TB by $32768^2$ pixels; beyond this its layer storage approaches node memory and \textsc{wsclean} switches to batched, disk-backed transforms \citep{Offringa2014}, avoiding crashes but inflating wall time, which is why it becomes impractical for the large-image runs of Section~\ref{sec:oskar_runtime}. \textsc{wgridder} is the most economical, applying a continuous-$w$ correction during gridding so that only a single image-sized Fourier grid is needed (the stricter $10^{-5}$ setting raises this by up to 50\%), while \textsc{idg} carries a high $\sim$20--25\,GB fixed floor but grows slowly thereafter, converging on \textsc{wgridder} by $32768^2$ pixels.

BRISK's footprint is dominated by the image-domain arrays, the grid-to-image FFT, and the compressed convolution functions; because it doesn't store full image-sized $w$-layers, it avoids \textsc{wstacking}'s steep growth, and the {\color{blue}interpolated BRISK} curve tracks the non-interpolated one closely, so the accuracy gains of Section~\ref{sec:oskar_accuracy} cost no memory. At $131072^2$ pixels it uses 790\,GB, of which the compressed kernels account for only 58\,GB. The limiting cost is therefore the image-sized arrays and FFT workspace, not the kernel representation: SVD compression shifts the wide-field memory bottleneck away from explicit $w$-layer storage and onto the image grid itself, giving a predictable image-bound scaling that reaches substantially larger images than the comparison methods on the same hardware.

\subsection{Performance on Realistic GLEAM Data}
\label{sec:gleam_results}

\subsubsection{Reconstruction Quality}
\label{sec:gleam_accuracy}

\begin{table*}[!b]
\centering
\captionsetup{font=small,skip=3pt}
\caption{Source-extraction and image-fidelity measurements on the GLEAM datasets. The number of true sources differs between image sizes because the field of view and catalogue-matching region increase with dataset scale. Matched entries give the number and percentage of catalogue sources recovered; false positives give unmatched detections; noise is reported in $\mu$Jy/beam.}
\label{tab:gleam_accuracy}
\footnotesize
\setlength{\tabcolsep}{3pt}
\resizebox{\textwidth}{!}{%
\begin{tabular}{l c c c c c c c}
\hline\hline
 & BRISK & BRISK + Interp. & WSTACKING & WSTACKING (PD) & WGRIDDER & WGRIDDER (PD) & CASA \\
\hline
\multicolumn{8}{l}{\textit{\textbf{Image Size 2048 (33 True Sources)}}} \\
\hline
Matched         & 29 (87.9\%) & 29 (87.9\%) & 29 (87.9\%) & 29 (87.9\%) & 29 (87.9\%) & 29 (87.9\%) & 29 (87.9\%) \\
False Positives & 1           & 7           & 4           & 4           & 6           & 6           & 7           \\
Noise ($\mu$Jy/b) & 737         & 737         & 947         & 965         & 765         & 793         & 679         \\
ITW-SSIM        & --          & 0.999       & 0.999       & 0.999       & 0.999       & 0.999       & 0.973       \\
\hline
\multicolumn{8}{l}{\textit{\textbf{Image Size 8192 (479 True Sources)}}} \\
\hline
Matched         & 389 (81.2\%) & 389 (81.2\%) & 389 (81.2\%) & 389 (81.2\%) & 389 (81.2\%) & 389 (81.2\%) & 326 (68.1\%) \\
False Positives & 1            & 1            & 1            & 1            & 1            & 1            & 68           \\
Noise ($\mu$Jy/b) & 743          & 740          & 751          & 727          & 703          & 678          & 2120         \\
ITW-SSIM        & --           & 0.999        & 0.999        & 0.999        & 0.999        & 0.999        & 0.983        \\
\hline
\multicolumn{8}{l}{\textit{\textbf{Image Size 16384 (1895 True Sources)}}} \\
\hline
Matched         & 900 (47.5\%) & {\color{blue}1043 (55.0\%)} & 803 (42.4\%) & 802 (42.3\%) & 928 (49.0\%) & 925 (48.8\%) & 701 (37.0\%) \\
False Positives & 0            & 0            & 0            & 0            & 0            & 0            & 0            \\
Noise ($\mu$Jy/b) & 194          & 164          & 282          & 283          & 172          & 173          & 407          \\
ITW-SSIM        & --           & 0.999        & 0.999        & 0.999        & 0.999        & 0.999        & 0.991        \\
\hline
\hline
\end{tabular}%
}
\end{table*}

\begin{figure*}[!t]
    \color{blue}
    \centering
    \includegraphics[width=0.97\textwidth]{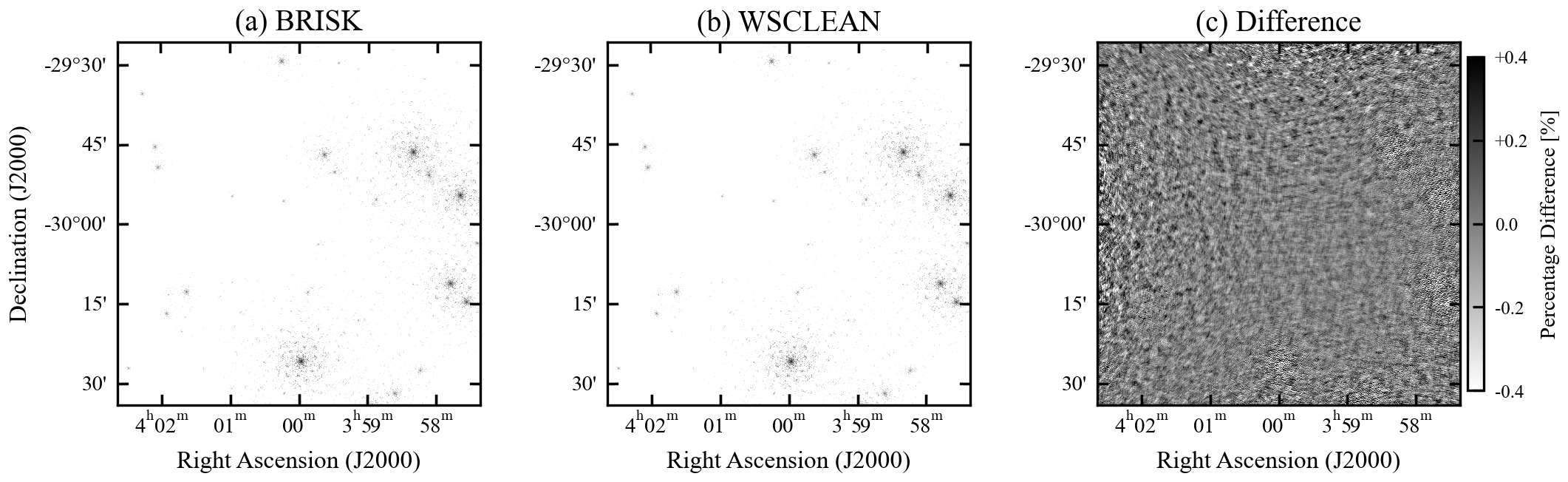}
    \caption{Dirty images of the small GLEAM dataset (Dataset~3, $2048^2$ pixels, $1\degr$ FoV centred on the EoR1 field) from (a) BRISK with interpolation and (b) \textsc{wsclean}'s \textsc{wgridder}, with (c) their difference in per cent of the dirty-image peak.}
    \label{fig:gleam_cleaned}
\end{figure*}

{\color{blue}
Figure~\ref{fig:gleam_cleaned} compares the BRISK and \textsc{wsclean} dirty images of the small GLEAM dataset directly: their difference has an rms of $0.13\%$ of the dirty-image peak, with the residual scatter well within $\pm0.4\%$ and dominated by sidelobe structure around the brightest sources. Table~\ref{tab:gleam_accuracy} summarises the source-extraction and image-fidelity statistics for all pipelines.
}
 
For the smallest dataset ($2048^2$ pixels, 33 catalogue sources), all pipelines recover the same 29 sources (87.9\%),  with each recording false positives and different measured noise. BRISK produces only one false positive — fewer than every other method, despite not having the lowest noise floor — indicating that at low source counts the BRISK reconstruction is conservative, recovering the dominant sources while introducing little spurious structure.
 
For the medium dataset ($8192^2$ pixels, 479 sources), BRISK and the \textsc{wsclean} gridders give near-identical statistics: each recovers 389 sources (81.2\%) with a single false positive and comparable noise floors. \textsc{casa} degrades clearly in this regime, recovering fewer sources (68.1\%), producing 68 false positives, and reaching a noise floor roughly three times higher. BRISK therefore matches the best \textsc{wsclean} configurations while avoiding the loss of completeness and rise in false detections seen for \textsc{casa}.
 
At the largest scale ($16384^2$ pixels, 1895 sources) recovery fractions fall for all methods, as the wider field holds far more faint sources near the detection threshold. {\color{blue}Here the interpolated BRISK recovers the most sources of any pipeline (55.0\%), ahead of \textsc{wgridder}, with the non-interpolated BRISK also outperforming both \textsc{wstacking} variants and \textsc{casa}.} It simultaneously reaches the lowest noise floor in the table (164\,$\mu$Jy/beam). All methods report zero false positives at this scale, so the matching threshold is set by sensitivity and completeness rather than spurious detections.
 
{\color{blue}
BRISK therefore maintains competitive, and at the largest scale leading, reconstruction quality on realistic data, with the interpolated variant again the more accurate, consistent with Section~\ref{sec:oskar_accuracy}.
}

\subsubsection{Runtime Scaling}
\label{sec:gleam_runtime}

\begin{figure}[!htbp]
    \centering
    \includegraphics[width=0.92\linewidth]{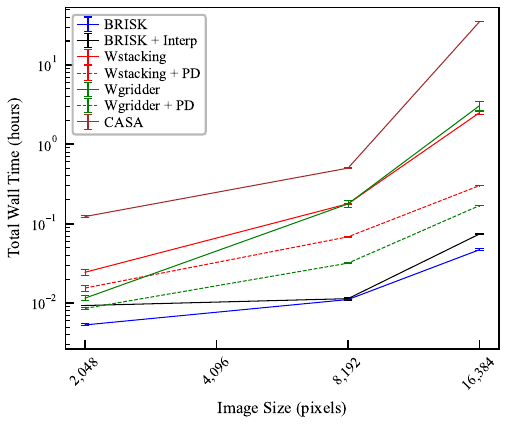}
    \caption{Wall time versus image size across the three SKAO GLEAM datasets (Table~\ref{tab:gleam_params}). Dashed curves denote \textsc{wsclean} runs with parallel deconvolution; solid curves denote standard CLEAN. \textsc{idg} is omitted because these datasets provide only the $XX$ polarisation.}
    \label{fig:gleam_runtime}
\end{figure}

{\color{blue}
Figure~\ref{fig:gleam_runtime} shows wall-time scaling for the GLEAM measurement sets, in which image size, field of view, observation duration and visibility count all increase together, giving a more representative end-to-end test than the controlled grid. 
}
 
BRISK achieves the lowest wall time at every configuration, and the separation widens with problem size. At the largest configuration ($16384^2$ pixels, $\sim$31 million visibilities) it completes in a few hundred seconds, while the \textsc{wsclean} gridders rise to several thousand seconds and \textsc{casa} becomes prohibitively expensive. The interpolated variant stays close to the non-interpolated runtime throughout, its overhead small relative to the gap to the comparison methods. Among those, \textsc{wgridder} is the closest baseline but both of its variants grow more steeply than BRISK; \textsc{wstacking} is slower again and scales less favourably, and \textsc{casa} is consistently slowest.

{\color{red}
Appendix~\ref{app:runtime_breakdown} (Table~\ref{tab:runtime_breakdown}) provides an end-to-end stage-by-stage profiling breakdown for BRISK, \textsc{wgridder (PD)}, and \textsc{casa} on the largest GLEAM dataset ($16384^2$ pixels). All runs are cold executions starting from un-cached Measurement Sets on disk, measuring data ingestion, kernel bank generation, FFT planning, and FITS output.

This reveals two primary bottlenecks in standard wide-field pipelines. First, in $w$-projection, \textsc{casa} requires 2,041.0\,s to compute its 2D convolution functions across 128 $w$-planes before any gridding. In contrast, BRISK generates its SVD-compressed 1D kernel bank across 268 $w$-planes in only 3.98\,s - a $513\times$ speedup in kernel generation alone. Second, during deconvolution, \textsc{casa}'s minor cycle dominates runtime (118,266\,s, for 435k components), whereas BRISK's tiled peak search and parallel minor-cycle execution complete 464k components in 47.90\,s. Together with forward/reverse gridding taking only 62.7\,s, BRISK finishes the complete cold run in 133.3\,s, achieving a $4.5\times$ speedup over \textsc{wsclean} (10.00\,min) and a $929\times$ speedup over \textsc{casa} (34.42\,hours).

The \textsc{casa} convolution-function build accounts for $1.6\%$ of its measured end-to-end runtime. In the optimistic limiting case, where a persistent cache removed the full 2,041\,s with no cache-read overhead, the \textsc{casa} total would fall from 123,913\,s to 121,872\,s (33.85\,h), still $914\times$ the measured cold-run BRISK wall time.
}
 
These results demonstrate that the performance advantage of the SVD-based approach persists under realistic observational conditions. The method remains fastest while also retaining competitive source-recovery statistics and image noise, as shown in Table~\ref{tab:gleam_accuracy}. This is important because the GLEAM datasets increase two computational drivers simultaneously: the number of visibilities and the image dimensions. BRISK therefore performs well not only on controlled source-grid benchmarks but also scales favourably on realistic survey data, making it well suited to large wide-field imaging problems where both image size and data volume drive computational cost.
 
\subsubsection{Peak Memory Usage}
\label{sec:gleam_memory}
\begin{figure}[!htbp]
    \centering
    \includegraphics[width=0.92\linewidth]{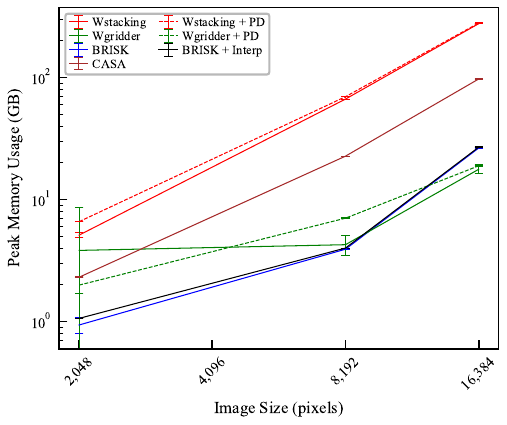}
    \caption{Peak memory usage versus image size across the three SKAO GLEAM datasets (Table~\ref{tab:gleam_params}). Dashed curves denote \textsc{wsclean} runs with parallel deconvolution; solid curves denote standard CLEAN.}
    \label{fig:gleam_memory}
\end{figure}

Figure~\ref{fig:gleam_memory} shows peak memory usage on the GLEAM datasets. The ordering mirrors the synthetic benchmarks of Section~\ref{sec:oskar_memory}, for the same mechanistic reasons. \textsc{wstacking} is again the most demanding, its stacked image-sized $w$-layers exceeding 280\,GB by $16384^2$ pixels. \textsc{casa} grows steeply too, reaching 100\,GB by $16384^2$ pixels, while \textsc{wgridder} — applying its $w$-correction continuously — remains the most economical, below 20\,GB even at the largest size in its PD configuration.

{\color{blue}
BRISK sits between \textsc{wgridder} and the heavier methods: a few GB at small sizes, growing smoothly to 27-28\,GB at $16384^2$ pixels, well below \textsc{wstacking} and \textsc{casa} and within roughly a factor of two of \textsc{wgridder}. As in the synthetic tests, interpolation adds only a modest overhead, so the improved source recovery and lower noise of Table~\ref{tab:gleam_accuracy} do not come at the cost of increased memory usage.
}

\section{Conclusions}
\label{sec:conclusion}

This work has presented BRISK, a wide-field radio interferometric imaging pipeline that replaces the large dense convolution-function libraries of conventional $w$-projection with a compact low-rank SVD representation, reconstructed on demand during gridding. {\color{blue}Where \citet{Merry2016} uses a single separable approximation and \citet{Pratley2019} accelerates construction of radially symmetric two-dimensional kernels, BRISK adapts the retained rank to each kernel to control low-rank reconstruction error while reducing storage and application cost.} Across both the controlled synthetic source grids and the realistic GLEAM datasets, this approach delivers reconstruction quality on par with the best available gridders while substantially reducing the cost of wide-field imaging.
 
On accuracy, the interpolated BRISK matches \textsc{wgridder} and \textsc{idg} in the large-image regime, reaching a peak-flux bias of {\color{blue}$-0.06\%$} and an ITW-SSIM of 0.997 at $32768^2$ pixels. On the realistic GLEAM data it goes further, achieving both the highest source-recovery fraction and the lowest image noise floor of any pipeline at the largest scale tested.
 
These fidelity levels are reached at a fraction of the cost. {\color{blue}BRISK records the lowest wall time of all tested pipelines, up to $103$--$132\times$ over the standard \textsc{wsclean} gridders, and a still-substantial $10$--$33\times$ even against \textsc{wsclean}'s parallel-deconvolution mode, at $32768^2$ pixels.} More importantly, BRISK is the only pipeline tested that remained stable to $131072^2$ pixels; comparison methods exceeded the runtime budget, exhausted memory, or diverged during deconvolution at substantially smaller sizes. {\color{blue}The parallel-deconvolution modes accelerate intermediate-scale runs but become unstable when cleaning to deep thresholds in large fields, whereas BRISK remains accurate and stable across the deep large-image runs tested.} This combination of speed, stability, and scalability, not faster gridding alone, is the principal result of this work.
 
By avoiding the wide kernel banks of traditional $w$-projection, BRISK remains competitive with memory-efficient continuous-$w$ methods such as \textsc{wgridder}, and at the largest scales its dominant cost is no longer convolution-function storage but the unavoidable image-sized arrays and FFT workspace. SVD compression therefore shifts the limiting bottleneck of wide-field imaging away from $w$-kernel storage and onto the cost of representing extremely large images themselves.

For all pipelines tested, including BRISK, the CLEAN minor-cycle subtraction and major-cycle corrections eventually dominate total runtime at sufficiently large image sizes. This suggests that future progress in wide-field imaging performance will depend increasingly on the development of scalable and stable deconvolution algorithms capable of operating efficiently on extremely large image planes.

{\color{red}
The present gridding/predict has addressed the $w$-term alone. Aperture ($A$-term) and other direction-dependent effects \citep{Bhatnagar2008} enter $w$-projection through the construction of the convolution functions rather than through the gridding step, and the on-the-fly reconstruction of Section~\ref{sec:kernel_recon} does not depend on how a kernel was generated. Extending the compressed kernel bank to $AW$-projection, where the number of stored kernels grows with antenna, time and frequency, is therefore a natural direction for future work. Likewise, wide-band sky modelling is outside the scope of the present narrow-band multi-frequency-synthesis implementation, and extending BRISK to incorporate frequency-dependent sky and primary-beam terms would be beneficial.

Furthermore, a GPU implementation of BRISK could provide further acceleration and portability for the next generation of telescopes. The SVD representation reduces the memory traffic of gridding rather than relying on any feature specific to CPUs, so the approach is not tied to the architecture on which it has been demonstrated here.
}

The current pipeline uses a conventional CLEAN framework; extending it with multi-scale CLEAN, adaptive thresholding, or optimisation-based deconvolution is another natural next step and is expected to improve reconstruction quality further, particularly for extended emission and highly structured diffuse sources. {\color{red}The combination of the additional development mentioned above would transform BRISK into an imaging pipeline capable of dealing with any measurement set, not only point source images. }

We also investigated whether similar SVD compression strategies could be applied to the point spread function. Unlike the gridding convolution functions, however, the PSF exhibits a complex sidelobe structure that does not admit aggressive low-rank truncation without substantial loss of fidelity. Consequently, PSF compression does not currently provide the same advantages as the gridding kernels.

Overall, this work demonstrates that low-rank SVD compression provides a viable, scalable, and highly effective solution for wide-field interferometric imaging. By achieving reconstruction quality comparable to and exceeding the most accurate existing gridders while substantially reducing runtime and controlling memory growth, BRISK enables imaging regimes that are difficult or impractical for existing approaches, making the method particularly well suited to the requirements of next-generation radio telescopes and SKA-scale imaging problems.

\section*{Data Availability}

The GLEAM-based SKAO simulation datasets used in this work are publicly available from the SKAO simulations repository:
\begin{center}
\url{https://gitlab.com/ska-telescope/sim/sim-datasets}
\end{center}

The synthetic SKA1-Mid benchmark measurement sets used for the controlled scaling experiments were generated specifically for this work using the OSKAR simulation package and the simulation parameters described in Section~\ref{sec:measurement_sets}.

The BRISK source code is available at reasonable request.



\bibliographystyle{plainnat}
\bibliography{svdw} 

@article{Scaife2020,
    author = {Scaife, A. M. M.},
    title = {Big telescope, big data: towards exascale with the Square Kilometre Array},
    journal = {Phil. Trans. R. Soc. A},
    fjournal = {Philosophical Transactions of the Royal Society A},
    volume = {378},
    number = {2166},
    pages = {20190060},
    year = {2020},
    doi = {10.1098/rsta.2019.0060},
}

@article{Braun2019,
    author = {Braun, R. and Bonaldi, A. and Bourke, T. and Keane, E. and Wagg, J.},
    title = {Anticipated Performance of the Square Kilometre Array -- Phase 1 (SKA1)},
    journal = {arXiv e-prints},
    year = {2019},
    eprint = {1912.12699},
    note = {arXiv:1912.12699},
}

@book{ThompsonMoranSwenson2017,
    author = {Thompson, A. R. and Moran, J. M. and Swenson, G. W.},
    title = {Interferometry and Synthesis in Radio Astronomy},
    edition = {3rd},
    publisher = {Springer Nature},
    year = {2017},
    doi = {10.1007/978-3-319-44431-4},
}

@inproceedings{Perley1999,
    author = {Perley, R. A.},
    title = {Imaging with Non-Coplanar Arrays},
    booktitle = {Synthesis Imaging in Radio Astronomy II},
    editor = {Taylor, G. B. and Carilli, C. L. and Perley, R. A.},
    series = {ASP Conf. Ser.},
    volume = {180},
    pages = {383},
    year = {1999},
}

@article{CornwellPerley1992,
    author = {Cornwell, T. J. and Perley, R. A.},
    title = {Radio-interferometric imaging of very large fields -- The problem of non-coplanar arrays},
    journal = {A\&A},
    fjournal = {Astronomy \& Astrophysics},
    volume = {261},
    pages = {353--364},
    year = {1992},
}

@article{Cornwell2008,
    author = {Cornwell, T. J. and Golap, K. and Bhatnagar, S.},
    title = {The Noncoplanar Baselines Effect in Radio Interferometry: The W-Projection Algorithm},
    journal = {IEEE J. Sel. Top. Signal Process.},
    fjournal = {IEEE Journal of Selected Topics in Signal Processing},
    volume = {2},
    number = {5},
    pages = {647--657},
    year = {2008},
    doi = {10.1109/JSTSP.2008.2005290},
}

@techreport{HumphreysCornwell2011,
    author = {Humphreys, B. and Cornwell, T. J.},
    title = {{SKA} Memo 132: Simulation of {SKA} Imaging Performance with the w-stacking Algorithm},
    institution = {SKA Office},
    year = {2011},
    note = {SKA Memo 132},
}

@inproceedings{CornwellVoronkovHumphreys2012,
    author = {Cornwell, T. J. and Voronkov, M. A. and Humphreys, B.},
    title = {Wide field imaging for the Square Kilometre Array},
    booktitle = {Image Reconstruction from Incomplete Data VII},
    series = {Proc. SPIE},
    volume = {8500},
    pages = {85000L},
    year = {2012},
    doi = {10.1117/12.929336},
}

@article{Offringa2019Precision,
    author = {Offringa, A. R. and Mertens, F. and van der Tol, S. and Veenboer, B. and Gehlot, B. K. and Koopmans, L. V. E. and Mevius, M.},
    title = {Precision requirements for interferometric gridding in the analysis of a 21 cm power spectrum},
    journal = {A\&A},
    volume = {631},
    pages = {A12},
    year = {2019},
    doi = {10.1051/0004-6361/201935722},
}

@article{Offringa2014,
    author = {Offringa, A. R. and McKinley, B. and Hurley-Walker, N. and Briggs, F. H. and Wayth, R. B. and Kaplan, D. L. and Bell, M. E. and Feng, L. and Neben, A. R. and Hughes, J. D. and others},
    title = {{WSClean}: an implementation of a fast, generic wide-field imager for radio astronomy},
    journal = {MNRAS},
    fjournal = {Monthly Notices of the Royal Astronomical Society},
    volume = {444},
    number = {1},
    pages = {606--619},
    year = {2014},
    doi = {10.1093/mnras/stu1368},
}

@misc{OffringaPDDocs,
    author = {{WSClean Developers}},
    title = {{WSClean} Parallel Deconvolution Documentation},
    year = {2024},
    note = {Accessed 2026; \url{https://wsclean.readthedocs.io/en/latest/parallel_deconvolution.html}},
}

@article{Arras2021,
    author = {Arras, P. and Reinecke, M. and Westermann, R. and En{\ss}lin, T. A.},
    title = {Efficient wide-field radio interferometry response},
    journal = {A\&A},
    fjournal = {Astronomy \& Astrophysics},
    volume = {646},
    pages = {A58},
    year = {2021},
    doi = {10.1051/0004-6361/202039723},
}

@article{Ye2022,
    author = {Ye, H. and Gull, S. F. and Tan, S. M. and Nikolic, B.},
    title = {High accuracy wide-field imaging method in radio interferometry},
    journal = {MNRAS},
    fjournal = {Monthly Notices of the Royal Astronomical Society},
    volume = {510},
    number = {3},
    pages = {4110--4125},
    year = {2022},
    doi = {10.1093/mnras/stab3548},
}

@article{vanderTol2018,
    author = {{van der Tol}, S. and Veenboer, B. and Offringa, A. R.},
    title = {Image Domain Gridding: a fast method for convolutional resampling of visibilities},
    journal = {A\&A},
    fjournal = {Astronomy \& Astrophysics},
    volume = {616},
    pages = {A27},
    year = {2018},
    doi = {10.1051/0004-6361/201832858},
}

@article{Merry2016,
    author = {Merry, B.},
    title = {Approximating {W} projection as a separable kernel},
    journal = {MNRAS},
    fjournal = {Monthly Notices of the Royal Astronomical Society},
    volume = {456},
    number = {2},
    pages = {1761--1766},
    year = {2016},
    doi = {10.1093/mnras/stv2761},
}

@article{Pratley2019,
    author = {Pratley, L. and Johnston-Hollitt, M. and McEwen, J. D.},
    title = {A Fast and Exact w-stacking and w-projection Hybrid Algorithm for Wide-field Interferometric Imaging},
    journal = {ApJ},
    fjournal = {The Astrophysical Journal},
    volume = {874},
    number = {2},
    pages = {174},
    year = {2019},
    doi = {10.3847/1538-4357/ab0a05},
}

@article{Pratley2020,
    author = {Pratley, L. and Johnston-Hollitt, M. and McEwen, J. D.},
    title = {w-Stacking w-projection hybrid algorithm for wide-field interferometric imaging: implementation details and improvements},
    journal = {Publ. Astron. Soc. Aust.},
    fjournal = {Publications of the Astronomical Society of Australia},
    volume = {36},
    pages = {e060},
    year = {2020},
    doi = {10.1017/pasa.2019.49},
}

@article{CASATeam2022,
    author = {{The CASA Team} and Bean, B. and Bhatnagar, S. and Castro, S. and Donovan Meyer, J. and Emonts, B. and others},
    title = {{CASA}, the Common Astronomy Software Applications for Radio Astronomy},
    journal = {PASP},
    fjournal = {Publications of the Astronomical Society of the Pacific},
    volume = {134},
    number = {1041},
    pages = {114501},
    year = {2022},
    doi = {10.1088/1538-3873/ac9642},
}

@article{Hogbom1974,
    author = {H{\"o}gbom, J. A.},
    title = {Aperture Synthesis with a Non-Regular Distribution of Interferometer Baselines},
    journal = {A\&AS},
    fjournal = {Astronomy \& Astrophysics Supplement Series},
    volume = {15},
    pages = {417--426},
    year = {1974},
}

@article{Schwab1984,
    author = {Schwab, F. R.},
    title = {Relaxing the isoplanatism assumption in self-calibration; applications to low-frequency radio interferometry},
    journal = {AJ},
    fjournal = {Astronomical Journal},
    volume = {89},
    pages = {1076--1081},
    year = {1984},
    doi = {10.1086/113605},
}

@phdthesis{Briggs1995,
    author = {Briggs, D. S.},
    title = {High fidelity deconvolution of moderately resolved sources},
    school = {The New Mexico Institute of Mining and Technology},
    year = {1995},
}

@article{Jackson1991,
    author = {Jackson, J. I. and Meyer, C. H. and Nishimura, D. G. and Macovski, A.},
    title = {Selection of a convolution function for Fourier inversion using gridding},
    journal = {IEEE Trans. Med. Imaging},
    fjournal = {IEEE Transactions on Medical Imaging},
    volume = {10},
    number = {3},
    pages = {473--478},
    year = {1991},
    doi = {10.1109/42.97598},
}

@article{HalkoMartinssonTropp2011,
    author = {Halko, N. and Martinsson, P.-G. and Tropp, J. A.},
    title = {Finding Structure with Randomness: Probabilistic Algorithms for Constructing Approximate Matrix Decompositions},
    journal = {SIAM Review},
    fjournal = {SIAM Review},
    volume = {53},
    number = {2},
    pages = {217--288},
    year = {2011},
    doi = {10.1137/090771806},
}

@article{FrigoJohnson2005,
    author = {Frigo, M. and Johnson, S. G.},
    title = {The Design and Implementation of {FFTW3}},
    journal = {Proc. IEEE},
    fjournal = {Proceedings of the IEEE},
    volume = {93},
    number = {2},
    pages = {216--231},
    year = {2005},
    doi = {10.1109/JPROC.2004.840301},
}

@article{Westmeier2021,
  author = {Westmeier, T. and Kitaeff, S. and Pallot, D. and others},
  title = {SoFiA 2 -- An automated, parallel HI source finding pipeline for the WALLABY survey},
  journal = {Monthly Notices of the Royal Astronomical Society},
  volume = {506},
  number = {3},
  pages = {3962--3978},
  year = {2021},
  doi = {10.1093/mnras/stab1881}
}

@inproceedings{LiArmour2022,
    author = {Li, X. and Armour, W.},
    title = {Intensity-Sensitive Similarity Indexes for Image Quality Assessment},
    booktitle = {2022 26th International Conference on Pattern Recognition (ICPR)},
    address = {Montreal, QC, Canada},
    pages = {1975--1981},
    year = {2022},
    doi = {10.1109/ICPR56361.2022.9956093},
}

@inproceedings{Mort2010,
    author = {Mort, B. J. and Dulwich, F. and Salvini, S. and Adami, K. Z. and Jones, M. E.},
    title = {{OSKAR}: Simulating digital beamforming for the {SKA} aperture array},
    booktitle = {2010 IEEE International Symposium on Phased Array Systems and Technology},
    pages = {690--694},
    year = {2010},
    doi = {10.1109/ARRAY.2010.5613289},
}

@article{vanHaarlem2013,
    author = {{van Haarlem}, M. P. and Wise, M. W. and Gunst, A. W. and Heald, G. and McKean, J. P. and Hessels, J. W. T. and de Bruyn, A. G. and Nijboer, R. and Swinbank, J. and Fallows, R. and others},
    title = {{LOFAR}: The {LOw-Frequency ARray}},
    journal = {A\&A},
    fjournal = {Astronomy \& Astrophysics},
    volume = {556},
    pages = {A2},
    year = {2013},
    doi = {10.1051/0004-6361/201220873},
}

@misc{SKALowSpecs,
  author = {{Square Kilometre Array Observatory}},
  title = {SKA Telescope Specifications},
  year = {2025},
  url = {https://www.skao.int/en/science-users/118/ska-telescope-specifications},
  note = {Accessed 2026-05-08}
}

@misc{SKAMidSpecs,
  author = {{Square Kilometre Array Observatory}},
  title = {SKA-Mid},
  year = {2025},
  url = {https://www.skao.int/en/explore/telescopes/ska-mid},
  note = {Accessed 2026-05-08}
}

@article{Bhatnagar2008,
    author = {Bhatnagar, S. and Cornwell, T. J. and Golap, K. and Us{\'o}n, J. M.},
    title = {Correcting direction-dependent gains in the deconvolution of radio interferometric images},
    journal = {A\&A},
    fjournal = {Astronomy \& Astrophysics},
    volume = {487},
    pages = {419--429},
    year = {2008},
    doi = {10.1051/0004-6361:20079284},
}

@techreport{Golap2015,
    author = {Golap, Kumar},
    title = {Multithreading Gridders Without Copying and Locks},
    institution = {National Radio Astronomy Observatory},
    number = {EVLA Memo 191},
    year = {2015},
    url = {https://library.nrao.edu/public/memos/evla/EVLAM_191.pdf},
}

@article{Veenboer2020,
    author = {Veenboer, Bram and Romein, John W.},
    title = {Radio-astronomical imaging on graphics processors},
    journal = {Astron. Comput.},
    fjournal = {Astronomy and Computing},
    volume = {32},
    pages = {100386},
    year = {2020},
    doi = {10.1016/j.ascom.2020.100386},
}

@techreport{Bhatnagar2022,
    author = {Bhatnagar, S. and Madsen, F. and Robnett, J.},
    title = {Baseline {HPG} runtime performance},
    institution = {National Radio Astronomy Observatory},
    number = {ngVLA Computing Memo 7},
    year = {2022},
    url = {https://library.nrao.edu/public/memos/ngvla/NGVLAC_07.pdf},
}

@article{Hodgson2026,
    author = {Hodgson, Torrance and Johnston-Hollitt, Melanie},
    title = {That pesky {A}-term: Efficiently correcting for direction-, time-, and baseline-dependent effects in radio interferometric imaging},
    journal = {Astron. Comput.},
    fjournal = {Astronomy and Computing},
    volume = {54},
    pages = {101012},
    year = {2026},
    doi = {10.1016/j.ascom.2025.101012},
}

\appendix
{\color{blue}
\section{BRISK implementation details}
\label{app:implementation}

This appendix records the standard imaging and implementation choices used in BRISK. These details are separated from the main description so that Section~\ref{sec:imager} can focus on the choices that directly affect the SVD-compressed kernel bank.

\begin{table*}
\color{blue}
  \centering
  \caption{\color{blue}Principal user-configurable imaging parameters. Defaults are shown in parentheses where applicable.}
  \label{tab:params}
  \begin{tabular}{lll}
    \hline
    Parameter & Role & Primary effect \\
    \hline
    \multicolumn{3}{l}{\textit{Image geometry}} \\
    Image size $W$ (pixels)      & Field of view        & Data volume and FoV \\
    Pixel scale $\delta\theta$   & Angular resolution   & Sampling of PSF \\
    $w$-planes $N_w$             & Non-coplanar correction & Accuracy vs.\ cost \\[4pt]
    \multicolumn{3}{l}{\textit{Visibility weighting}} \\
    Weighting scheme             & PSF shape            & Resolution-sensitivity trade-off \\
    Robust parameter $R_{\rm robust}$ ($0$) & Briggs interpolation & $-2$: uniform; $+2$: natural \\[4pt]
    \multicolumn{3}{l}{\textit{Kernel accuracy}} \\
    AA support $S_{\rm AA}$ (7\,pix) & Alias suppression & Wider: lower aliasing, higher cost \\
    Kaiser $\beta$ (9.0)         & Sidelobe level       & Larger: lower sidelobes, broader lobe \\
    Support threshold $\kappa$ ($1-10^{-6}$) & $w$-kernel spatial support & Lower: smaller kernels, faster gridding \\
    Rank threshold $\epsilon$ ($1-10^{-6}$) & Retained SVD rank $r_{\rm SVD}$ & Lower: fewer modes, faster gridding \\[4pt]
    \multicolumn{3}{l}{\textit{Deconvolution}} \\
    Loop gain $\gamma$ (0.1)     & Per-iteration subtraction & Smaller: slower, more stable \\
    Major-cycle fraction $\gamma_{\rm major}$ & Minor-cycle depth & Controls major-cycle frequency \\
    Threshold $n_\sigma$ (3)     & Stopping depth       & Sets deconvolution floor \\[4pt]
    \multicolumn{3}{l}{\textit{Computational}} \\
    Oversampling $N_{\rm os}$ (63) & Sub-pixel accuracy & Max position error $1/(2N_{\rm os})$ cell \\
    FFT padding $f_{\rm pad}$ (1.2) & Aliasing suppression & Larger: cleaner edges, larger FFT \\
    Thread count                 & Parallelism          & Scales gridding and CLEAN throughput \\
    \hline
\end{tabular}
\end{table*}

\subsection{Visibility data and weighting}
\label{sec:weighting}
Each visibility is handled by the gridder as a weighted $(u,v,w)$ tuple after its coordinates are converted to wavelengths at its own channel frequency; no per-channel bookkeeping, spectral-index correction or Taylor-term expansion is applied. Natural, uniform and Briggs weighting are available. Uniform weighting divides each weight by the number of samples in its $uv$ cell of size $1/(W\,\delta\theta)$, where $W$ is the image size in pixels and $\delta\theta$ the pixel scale, and Briggs weighting \citep{Briggs1995} interpolates between the two through the robust parameter $R_{\rm robust}$, approaching uniform at $-2$ and natural at $+2$. Density cells are defined on the zero-padded grid so that the weighting resolution matches the image scale.

\subsection{Anti-aliasing kernel}
\label{sec:AA_kernel}
The anti-aliasing (AA) function is a separable taper that is used to suppress the aliasing that arises from truncating the continuous $uv$-domain onto a discrete grid. As mentioned in Section~\ref{sec:wproj}, the AA kernel is incorporated into the GCF before gridding. It is defined as the product of a one-dimensional window function with itself, and in our imager, it has three kernel options dependent on the user's preference.

The default is the Kaiser-Bessel window (as used in \textsc{wsclean}),
\begin{equation}
k(d) = \frac{I_0\!\left(\beta\sqrt{1-(d/a)^2}\right)}{I_0(\beta)},
\qquad |d| \le a,
\end{equation}
where $d$ is the pixel offset from the kernel centre, $a$ is the half-support, $I_0$ is the zeroth-order modified Bessel function of the first kind, and $\beta$ is a shape parameter that controls the trade-off between the width of the main lobe and the level of aliasing sidelobes. \citet{Offringa2014} found that the optimum support width for this function in terms of an accuracy and run time trade-off is 7, so this is the default value in our imager. This can be increased if the user desires higher image accuracy, often at the cost of runtime.

Other options include the sinc-Kaiser variant, which multiplies the Kaiser envelope by a sinc function to prevent the kernel from reaching 0 at the support boundary, and the prolate spheroidal wave function (PSWF) (as used in \textsc{casa}), which is the theoretically optimal bandlimiting window as it provides the lowest possible aliasing level for a given kernel size at the expense of increased compute cost.

The image-domain AA taper is formed by taking the iFFT of the padded AA kernel in the $uv$-domain. This is used during kernel construction for the GCF and then again to correct for the apodisation introduced during gridding through a per-pixel division of the dirty image and PSF.

\subsection{$w$-kernel construction and support selection}
\label{sec:wkern}

\subsubsection{$w$-plane discretisation}
\label{sec:wplane_discretisation}

The number of stored planes $N_w$ is a user parameter (Section~\ref{sec:param_validation}); increasing it reduces the $w$-discretisation error at the cost of constructing and storing more compressed kernels, and the adjacent-plane interpolation of Section~\ref{sec:wplane_interpolation} reduces that error further for a given $N_w$. The $\sqrt{|w|}$ plane distribution of Section~\ref{sec:imager} places more planes near $w=0$ and samples the high-$|w|$ regime more coarsely. For the arrays considered here the visibility density falls steeply with $|w|$, since large $|w|$ arises only from the longest projected baselines, so concentrating planes at low $|w|$ reduces the count-weighted nearest-plane error while the sparsely populated high-$|w|$ regime tolerates coarser sampling.

The geometric cap $w_{\max,{\rm geom}} = 1/(4|\Delta l|)$ applied to $w_{\max}$ in Section~\ref{sec:imager} reflects the fact that, for a pixel scale of $|\Delta l|$ radians, the $w$-projection phase screen contains structure finer than the image sampling once $|w|$ exceeds this value, so higher-$w$ kernels would represent oscillations that the image grid cannot resolve. The cap prevents the bank from constructing unnecessarily large supports or allocating planes to unrepresentable phase structure, which matters most when imaging coarse fields that contain very long-baseline data.

\subsubsection{Energy-based support selection}
To select the support of the GCF kernels, the radially averaged energy profile in the $uv$-domain is computed and accumulated, starting at the centre and moving outward. The kernel is then cropped to the smallest radius $S$, corresponding to a full width $N = 2S+1$ as defined in Section~\ref{sec:SVDkernels}, such that the enclosed energy fraction satisfies
\begin{equation}
\frac{E_{\rm enclosed}(S)}{E_{\rm total}} \ge \kappa,
\end{equation}
where $\kappa$ is a user input (set to $0.999999$ by default) depending on their desired trade-off between image accuracy and runtime. This support selection is done for each $w$-plane GCF independently so that the large $|w|$ kernels receive a larger support. The cropped kernel is then SVD-compressed as described in Section~\ref{sec:SVDkernels}, where the separate rank threshold $\epsilon$ sets how many singular modes are retained.

\subsection{Dirty image and PSF formation}
\label{sec:dirty}
All visibilities are gridded onto a zero-padded grid to suppress circular aliasing during the FFT. After the FFT, the central unpadded image patch is extracted and de-apodised. The PSF is formed via an identical pass using the same weights but with unit visibility amplitudes. After de-apodisation, the PSF is normalised so that the central pixel equals one. This also establishes the flux scaling for the dirty image and subsequent deconvolution.

The Fourier transforms make use of the FFTW library \citep{FrigoJohnson2005} with the \texttt{MEASURE} planning strategy to generate optimised execution plans. These plans are saved to a wisdom file that can be used in subsequent runs, so the one-time planning overhead is incurred only once per CPU node and image size. A warm-up mode has been provided to pre-build plans before any full deconvolution run.

\subsection{Deconvolution}
\label{sec:clean}

\subsubsection{Noise estimation and stopping threshold}
Before each major cycle, the noise level in the residual image is estimated using the median absolute deviation (MAD),
\begin{equation}
\hat{\sigma} = 1.4826 \times
  \mathrm{median}\!\left(\left|I^{\rm res}_i - \widetilde{I}^{\rm res}\right|\right),
\end{equation}
where $\widetilde{I}^{\rm res}$ is the image median. The scaling factor $1.4826 = \Phi^{-1}(0.75)^{-1}$ makes the estimator consistent with the standard deviation of a Gaussian distribution. The noise estimate is recalculated at the start of every major cycle so that the threshold becomes more accurate as the residual improves.

The deconvolution cycles continue until a stopping criterion is reached. In automatic mode, the threshold is set to $n_\sigma\,\hat{\sigma}$ with $n_\sigma = 3$ by default, and the pipeline continues until the peak residual falls below this level. In manual mode, the user specifies the desired threshold. This is appropriate when the noise behaviour is known in advance or when a specific depth of cleaning is required.

\subsubsection{CLEAN cycles}
The minor cycle makes use of H\"ogbom CLEAN. A PSF patch is selected by taking a fixed fraction of the PSF energy. At each iteration, the global peak of the residual is selected, and a fraction $\gamma$ (the loop gain, default $\gamma=0.1$) of that peak is added to the model. The PSF patch, scaled by the peak and loop gain, is then subtracted from the residual image at the location of the peak. This is repeated until the peak falls below either the stopping threshold or, in Cotton--Schwab mode, $(1-\gamma_{\rm major})$ times the initial peak in the current minor cycle. The parameter $\gamma_{\rm major}$ controls how deeply the minor cycle cleans before triggering a major cycle, in which visibilities are predicted from the current sky model using the full measurement operator and an updated residual image is formed from the data.

To avoid rescanning the entire image after each subtraction, the residual image is partitioned into tiles. Each tile maintains its current maximum and is marked if the PSF subtraction footprint overlaps it. In the next peak-finding step, only those tiles are rescanned. This reduces the per-iteration peak-search cost from $\mathcal{O}(W^2)$ to $\mathcal{O}(N_{\rm invalid})$ when the PSF support is compact relative to the image. The global maximum remains exact at every iteration.

On exiting a minor cycle, the model image is Fourier transformed to the $uv$ domain and passed through the prediction step described in Section~\ref{sec:predict}. The residual visibilities are updated as $V^{\rm res}=V^{\rm data}-V^{\rm model}$ so that $w$-projection errors are corrected at each major cycle rather than accumulated during deconvolution.

\subsubsection{Restoring beam}
After the stopping threshold has been reached, the restoring beam is estimated by fitting an elliptical Gaussian to the main lobe of the PSF. The CLEAN model is convolved with the beam and added to the final residual image to form the restored image. Users may override the fitted beam with explicit values for the major and minor axes and position angle.

\enlargethispage{2\baselineskip}
\subsection{Imaging parameters}
\label{sec:params}

Table~\ref{tab:params} lists the parameters most likely to require adjustment for a given observation. All remain user-configurable so that the pipeline can be adapted to the required accuracy and runtime.
}

\clearpage
\onecolumn
{\color{blue}
\section{Measurement-set parameters}
\label{app:measurement_sets}

\begin{center}
    \captionsetup{font=footnotesize,skip=3pt,width=0.82\textwidth}
    \small
    \setlength{\tabcolsep}{3pt}
    \captionof{table}{\color{blue}Simulation parameters for the two OSKAR point-source datasets (Datasets~1 and~2). Both datasets share the same SKA1-Mid telescope model and observing geometry; they differ only in their source configuration and their spectral and temporal sampling.}
    \label{tab:oskar_params}
    \begin{tabular}{lrr}
        \hline
        Parameter                   & Dataset~1                       & Dataset~2                   \\
        \hline
        Array                       & SKA1-Mid AA2                    & SKA1-Mid AA2                \\
        Number of stations          & 63                              & 63                          \\
        Number of baselines         & 1953                            & 1953                        \\
        Station beam                & Isotropic                       & Isotropic                   \\
        Polarisation mode           & Full                            & Full                        \\
        Phase centre (RA, Dec)      & $(0.2\degr, -30.72\degr)$       & $(0.2\degr, -30.72\degr)$   \\
        Field of view               & $8\degr$                        & $8\degr$                    \\
        Source configuration        & $15\times15$ grid (225 sources) & 50 random sources           \\
        Start frequency             & 1.2\,GHz                        & 1.2\,GHz                    \\
        Number of channels          & 1                               & 50                          \\
        Channel bandwidth           & 2\,MHz                          & 2\,MHz                      \\
        Total bandwidth             & 2\,MHz                          & 100\,MHz                    \\
        Observation length          & 2\,h                            & 2\,h                        \\
        Number of time steps        & 200                             & 800                         \\
        Time resolution             & 36\,s                           & 9\,s                        \\
        Time average per dump       & --                              & 2\,s                        \\
        Total visibilities          & 390\,600                        & 78\,120\,000                \\
        Ionosphere                  & None                            & None                        \\
        Direction-dependent effects & Disabled                        & Disabled                    \\
        \hline
    \end{tabular}
    \vspace{1.2em}

    \captionof{table}{\color{blue}Simulation parameters for the SKAO GLEAM datasets
             (Dataset~3). Image resolution is kept constant across
             all three sizes at $2$\,arcsec\,pixel$^{-1}$; visibility counts include all baselines, time steps, and frequency channels for each configuration.}
    \label{tab:gleam_params}
    \begin{tabular}{lrrr}
        \hline
        Parameter & Small & Medium & Large \\
        \hline
        Array                       & SKA1-Low      & SKA1-Low      & SKA1-Low      \\
        Number of stations          & 512           & 512           & 512           \\
        Number of baselines         & 130\,816      & 130\,816      & 130\,816      \\
        Sky model                   & GLEAM EoR1    & GLEAM EoR1    & GLEAM EoR1    \\
        Number of sources           & 33            & 479           & 1895        \\
        Phase centre (RA, Dec)      & $(60\degr, -30\degr)$ & $(60\degr, -30\degr)$ & $(60\degr, -30\degr)$ \\
        Field of view               & $1\degr$      & $4\degr$      & $8\degr$      \\
        Image size                  & $2048\times2048$  & $8192\times8192$  & $16384\times16384$ \\
        Pixel size (arcsec)         & 2             & 2             & 2            \\
        Observation duration        & 15\,min       & 1\,h          & 4\,h          \\
        Number of time steps        & 30            & 30            & 60            \\
        Time resolution             & 30\,s         & 2\,min        & 4\,min        \\
        Number of channels          & 1             & 2             & 4             \\
        Total visibilities          & 3\,924\,480   & 7\,848\,960   & 31\,395\,840  \\
        \hline
    \end{tabular}
\end{center}
}

\clearpage
\section{15 by 15 grid accuracy table}
\label{app:gridder_accuracy}
\begin{center}
\footnotesize
\captionof{table}{Source-extraction and image-fidelity statistics as a function of image size. Peak-flux entries give the mean recovered source peak in Jy/beam, with the minimum and maximum recovered values in square brackets. PD denotes the \textsc{wsclean} parallel-deconvolution mode.}
\label{tab:gridder_accuracy}
\begin{tabular}{l l c c c c}
\hline\hline
Image Size & Gridder & Mean Peak Flux (Jy/b) & Std. Deviation (Jy/b) & Noise Floor (mJy/b) & ITW-SSIM \\
\hline
1024 & BRISK & {\color{blue}1.0616 [0.7977, 1.2591]} & {\color{blue}0.0829} & {\color{blue}21.33} & -- \\
 & BRISK + Interp. & {\color{blue}1.0624 [0.8345, 1.2563]} & {\color{blue}0.0834} & {\color{blue}21.30} & 0.999 \\
 & WSTACKING & 0.7869 [0.5574, 1.0312] & 0.1168 & 21.57 & 0.998 \\
 & WSTACKING (PD) & 0.7868 [0.5401, 1.0311] & 0.1193 & 21.17 & 0.998 \\
 & WGRIDDER & 0.7922 [0.5752, 1.1658] & 0.1197 & 21.07 & 0.997 \\
 & WGRIDDER (PD) & 0.7896 [0.5753, 1.0282] & 0.1184 & 21.07 & 0.997 \\
 & IDG & 0.8194 [0.5524, 1.2047] & 0.1288 & 20.43 & 0.995 \\
 & IDG (PD) & 0.8194 [0.5524, 1.2047] & 0.1288 & 20.43 & 0.995 \\
 & CASA & 1.0243 [0.8040, 1.1460] & 0.0697 & 9.27 & 0.995 \\
\hline
2048 & BRISK & {\color{blue}0.9730 [0.8123, 1.0385]} & {\color{blue}0.0481} & {\color{blue}6.83} & -- \\
 & BRISK + Interp. & {\color{blue}0.9737 [0.8100, 1.0368]} & {\color{blue}0.0504} & 6.70 & 0.999 \\
 & WSTACKING & 0.9539 [0.7549, 1.0583] & 0.0657 & 7.00 & 0.999 \\
 & WSTACKING (PD) & 0.9530 [0.7582, 1.0532] & 0.0648 & 6.79 & 0.999 \\
 & WGRIDDER & 0.9432 [0.7306, 1.0392] & 0.0721 & 4.31 & 0.999 \\
 & WGRIDDER (PD) & 0.9531 [0.7475, 1.0475] & 0.0679 & 6.48 & 0.999 \\
 & IDG & 0.9394 [0.7282, 1.0367] & 0.0724 & 4.04 & 0.998 \\
 & IDG (PD) & 0.9383 [0.7257, 1.0362] & 0.0727 & 3.99 & 0.998 \\
 & CASA & 0.9811 [0.8636, 1.0165] & 0.0276 & 4.90 & 0.996 \\
\hline
4096 & BRISK & {\color{blue}0.9872 [0.9682, 0.9996]} & {\color{blue}0.0062} & {\color{blue}4.43} & -- \\
 & BRISK + Interp. & {\color{blue}1.0010 [0.9648, 1.0241]} & {\color{blue}0.0078} & {\color{blue}2.23} & 0.998 \\
 & WSTACKING & 0.9845 [0.9337, 1.0031] & 0.0167 & 3.51 & 0.999 \\
 & WSTACKING (PD) & 0.9800 [0.9272, 1.0004] & 0.0166 & 4.62 & 0.999 \\
 & WGRIDDER & 0.9919 [0.9238, 1.0019] & 0.0150 & 1.64 & 0.995 \\
 & WGRIDDER (PD) & 0.9883 [0.9191, 0.9988] & 0.0156 & 1.36 & 0.996 \\
 & IDG & 0.9919 [0.9223, 1.0015] & 0.0151 & 1.64 & 0.995 \\
 & IDG (PD) & 0.9883 [0.9177, 0.9988] & 0.0158 & 1.35 & 0.996 \\
 & CASA & 0.9858 [0.9239, 1.0138] & 0.0166 & 3.80 & 0.978 \\
\hline
8192 & BRISK & {\color{blue}0.9730 [0.9358, 0.9874]} & {\color{blue}0.0096} & 4.20 & -- \\
 & BRISK + Interp. & {\color{blue}1.0058 [0.9877, 1.0445]} & {\color{blue}0.0114} & {\color{blue}0.60} & 0.992 \\
 & WSTACKING & 0.9665 [0.8802, 0.9995] & 0.0262 & 5.16 & 0.996 \\
 & WSTACKING (PD) & 0.9645 [0.8808, 0.9977] & 0.0257 & 5.18 & 0.996 \\
 & WGRIDDER & 0.9980 [0.9937, 1.0014] & 0.0013 & 0.79 & 0.991 \\
 & WGRIDDER (PD) & 0.9970 [0.9925, 1.0018] & 0.0020 & 0.56 & 0.993 \\
 & IDG & 0.9981 [0.9939, 1.0016] & 0.0012 & 0.79 & 0.991 \\
 & IDG (PD) & 0.9970 [0.9924, 1.0021] & 0.0020 & 0.56 & 0.993 \\
 & CASA & 0.9717 [0.8947, 1.0116] & 0.0209 & 2.03 & 0.843 \\
\hline
16384 & BRISK & {\color{blue}0.9948 [0.9735, 1.0090]} & {\color{blue}0.0065} & {\color{blue}6.10} & -- \\
 & BRISK + Interp. & {\color{blue}1.0014 [0.9878, 1.0207]} & {\color{blue}0.0061} & {\color{blue}2.73} & 0.994 \\
 & WSTACKING & 0.9052 [0.6678, 0.9883] & 0.0677 & 8.75 & 0.990 \\
 & WSTACKING (PD) & 0.9061 [0.6666, 1.0005] & 0.0681 & 8.66 & 0.990 \\
 & WGRIDDER & 0.9826 [0.9721, 0.9938] & 0.0046 & 1.14 & 0.994 \\
 & WGRIDDER (PD) & 0.9721 [0.9469, 1.0076] & 0.0121 & 2.50 & 0.987 \\
 & IDG & 0.9836 [0.9735, 0.9988] & 0.0045 & 1.20 & 0.995 \\
 & IDG (PD) & 0.9733 [0.9356, 1.0154] & 0.0147 & 2.90 & 0.985 \\
\hline
32768 & BRISK & {\color{blue}0.9918 [0.9729, 1.0098]} & {\color{blue}0.0066} & {\color{blue}4.97} & -- \\
 & BRISK + Interp. & {\color{blue}0.9994 [0.9845, 1.0177]} & {\color{blue}0.0053} & {\color{blue}2.03} & 0.997 \\
 & WSTACKING & 0.8294 [0.4513, 1.0272] & 0.1528 & 7.59 & 0.923 \\
 & WSTACKING (PD) & 0.8530 [0.5120, 1.0558] & 0.1527 & 6.77 & 0.929 \\
 & WGRIDDER & 0.9945 [0.9874, 1.0038] & 0.0028 & 1.25 & 0.996 \\
 & WGRIDDER (PD) & 1.0428 [0.9044, 1.1748] & 0.0456 & 3.24 & 0.891 \\
 & IDG & 0.9943 [0.9865, 1.0049] & 0.0028 & 1.26 & 0.996 \\
 & IDG (PD) & 1.0081 [0.7455, 1.2191] & 0.0858 & 1.99 & 0.856 \\
\hline
65536 & BRISK & {\color{blue}1.0039 [0.9851, 1.0157]} & {\color{blue}0.0052} & {\color{blue}2.42} & -- \\
 & BRISK + Interp. & {\color{blue}0.9995 [0.9883, 1.0108]} & {\color{blue}0.0032} & {\color{blue}1.50} & 0.999 \\
\hline
131072 & BRISK & {\color{blue}0.9861 [0.9698, 1.0036]} & {\color{blue}0.0059} & {\color{blue}2.24} & -- \\
 & BRISK + Interp. & {\color{blue}1.0050 [0.9952, 1.0177]} & {\color{blue}0.0038} & {\color{blue}0.52} & 0.999 \\
\hline
\end{tabular}
\end{center}
\FloatBarrier

\clearpage
{\color{red}
\section{CPU and GPU runtime breakdown on the 50-channel Dataset~2}
\label{app:idg_gpu_breakdown}

\begin{center}
\small
\captionof{table}{\color{red}Stage-by-stage wall-time comparison of dataset 2 at $16384^2$ pixels. All runs used 48 cores of an Intel Xeon Gold 6548N processor; the GPU and hybrid modes additionally used one 48-GB NVIDIA L40S. Timings are in seconds, with the percentage of total application wall time in parentheses.}
\label{tab:idg_gpu_breakdown}
\setlength{\tabcolsep}{5pt}
\begin{tabular}{lrrrr}
\hline\hline
Pipeline Stage & BRISK & IDG CPU & IDG GPU & IDG Hybrid \\
\hline
Data Ingestion \& Weighting    & 3.68\,s (1.5\%)    & --                     & --                     & --                     \\
Kernel Bank Generation         & 10.08\,s (4.1\%)   & --                     & --                     & --                     \\
Inversion (Gridding \& FFT)    & 84.32\,s (33.9\%)  & 766.62\,s (25.6\%)    & 208.37\,s (9.7\%)     & 207.21\,s (9.5\%)     \\
Prediction (Degridding \& FFT) & 104.84\,s (42.2\%) & 380.21\,s (12.7\%)    & 145.30\,s (6.7\%)     & 143.91\,s (6.6\%)     \\
Minor-Cycle Deconvolution      & 29.01\,s (11.7\%)  & 1821.87\,s (60.8\%)   & 1780.95\,s (82.6\%)   & 1797.89\,s (82.9\%)   \\
Restoration \& Output          & 10.81\,s (4.4\%)   & --                     & --                     & --                     \\
Other / Overhead               & 5.74\,s (2.3\%)    & 25.60\,s (0.9\%)      & 21.11\,s (1.0\%)      & 20.89\,s (1.0\%)      \\
\hline
Total Wall Time                & 248.48\,s & 2994.30\,s & 2155.73\,s & 2169.90\,s \\
Slowdown vs BRISK              & --                    & 12.05$\times$            & 8.68$\times$             & 8.73$\times$             \\
\hline\hline
\end{tabular}
\end{center}
}

{\color{red}
\section{Runtime breakdown on realistic GLEAM data}
\label{app:runtime_breakdown}

\begin{center}
\small
\captionof{table}{\color{red}Stage-by-stage wall-time breakdown for cold runs on the large GLEAM dataset ($16384^2$ pixels). Timings are reported in seconds, with the percentage of total wall time shown in parentheses.}
\label{tab:runtime_breakdown}
\setlength{\tabcolsep}{6pt}
\begin{tabular}{lccc}
\hline\hline
Pipeline Stage & BRISK & WGRIDDER (PD) & CASA \\
\hline
Data Ingestion \& Weighting         & 4.01\,s (3.0\%)                & 14.40\,s (2.4\%)                & 16.00\,s (0.01\%) \\
Kernel Bank Generation              & 3.98\,s (3.0\%)       & 0.00\,s (N/A)                   & 2041.00\,s (1.6\%) [34.0\,min] \\
PSF, Beam \& Setup                  & 2.86\,s (2.1\%)                & --                               & 202.00\,s (0.2\%) [3.4\,min] \\
Inversion (Gridding \& FFT)         & 28.46\,s (21.3\%)              & 184.10\,s (30.7\%)              & 1540.00\,s (1.2\%) [25.7\,min] \\
Prediction (Degridding \& FFT)      & 34.25\,s (25.7\%)              & 133.81\,s (22.3\%)              & 1751.00\,s (1.4\%) [29.2\,min] \\
Minor-Cycle Deconvolution           & 47.90\,s (35.9\%)              & 251.11\,s (41.8\%)              & 118266.00\,s (95.4\%) [32.8\,hr] \\
Restoration \& Output               & 11.63\,s (8.7\%)               & 16.83\,s (2.8\%)                & 97.00\,s (0.1\%) [1.6\,min] \\
Other / Overhead                    & 0.25\,s (0.2\%)                & 0.00\,s (0.0\%)                 & 0.00\,s (0.0\%) \\
\hline
Total Wall Time           & 133.34\,s (2.22\,min) & 600.25\,s (10.00\,min) & 123913.00\,s (34.42\,hr) \\
Slowdown vs BRISK          & --          & 4.50$\times$   & 929.30$\times$ \\
\hline\hline
\end{tabular}
\end{center}
}

\end{document}